%% file: EPTTestQED.tex
\documentclass{article}
\usepackage{amsmath,amsbsy,amsfonts}
\usepackage{graphics}
\usepackage{graphicx}

\begin{document}
\input{Commands}
\input{FigurecommandsV.tex}
\input{FigGreen}
\setlength{\unitlength}{0.6cm}
\renewcommand{\Ga}{\Gamma}
\newcommand{\GP}{\Gamma_P}
\newcommand{\GPP}{\Gamma_P}

\title{Energy-dependent perturbation theory:\\ \large Possibility for improved tests of quantum-electrodynamics}

\author{Ingvar Lindgren$^1$, Sten Salomonson$^1$, and Johan Holmberg$^{1,2}$\\
$^1$ Department of Physics, University of Gothenburg\\
Gothenburg, Sweden\\
$^2$ Department of Physics, University of Heidelberg, Germany}

\maketitle

\begin{abstract}
Measurements of energy separations in highly charged ions can in many cases nowadays be performed with very high accuracy --- an accuracy that sometimes cannot be matched by the corresponding theoretical calculations. Furthermore, it has recently been demonstrated (Chantler \it{\it{et al.}} Phys. Rev. Lett. \textbf{109}, 153001, 2012) that there is a systematic deviation between experimental and theoretical results for the K$_\alpha$ X-ray lines of medium-heavy heliumlike ions. We have during a number of years been developing a general procedure for energy-dependent perturbative calculations, which opens up a unique possibility of incorporating the energy-dependent QED perturbations into the all-order many-body perturbation expansion in a rigorous way. Such an expansion will yield several important effects, never before accounted for in this type of analysis, which is expected to increase the theoretical accuracy considerably.  Calculation of some of these effects have been performed at our laboratory in Gothenburg, and numerical results are given. Further work along this line is now in progress. To what extent the improved procedure might explain the discrepancy found by Chantler \it{\it{et al.}} remains to be seen.
\end{abstract}

\newcommand{\contract}[3]
{\setlength{\unitlength}{1cm}
\begin{picture}(0,0.5)(#2,#3)
\thinlines\put(0,0){\line(1,0){#1}}
\put(0,0){\line(0,-1){0.2}}\put(#1,0){\line(0,-1){0.2}}
\end{picture}}

\setlength{\unitlength}{0.7cm}

\section{Introduction}
The theory of quantum electrodynamics (QED) has been very well tested for simple systems, like the free electron (anomalous magnetic moment) or the hydrogen atom (Lamb shift). For multi-electron systems, on the other hand, the  corresponding tests are much less complete. There is presently a great research interest in testing QED, particularly of highly charged ions, which can be produced in good intensities at large accelerators, like that at the GSI facility in Darmstadt.

In recent years very accurate experimental data have been produced for a number of highly charged few-electron ions, as well as for some inner-shell transitions of many-electron systems. The corresponding theoretical analysis, however, can in some cases not match the experimental accuracy. In Tables (\ref{Tab:Si}) and (\ref{Tab:CuXray}) we show two examples of this situation, a transition in He-like silicon, measured with laser techniques by Myers \it{et al.}~\cite{Myers08}, and the K$_\alpha$ X-ray lines in the copper atom, studied by Deslattes \it{et al.}~\cite{Deslattes03}  The theoretical estimates are made by means of two-photon exchange (Artemyev~\cite{AShab05}) or relativistic many-body technique with analytical QED energies added (Plante~\cite{PJS94}, Chantler/Grant~\cite{Chantler10}). The experimental results are here 1-2 orders of magnitude more accurate.

\begin{figure}[h]
\begin{picture}(10,12)(-0.5,0.3)
{\includegraphics[scale=1.3]{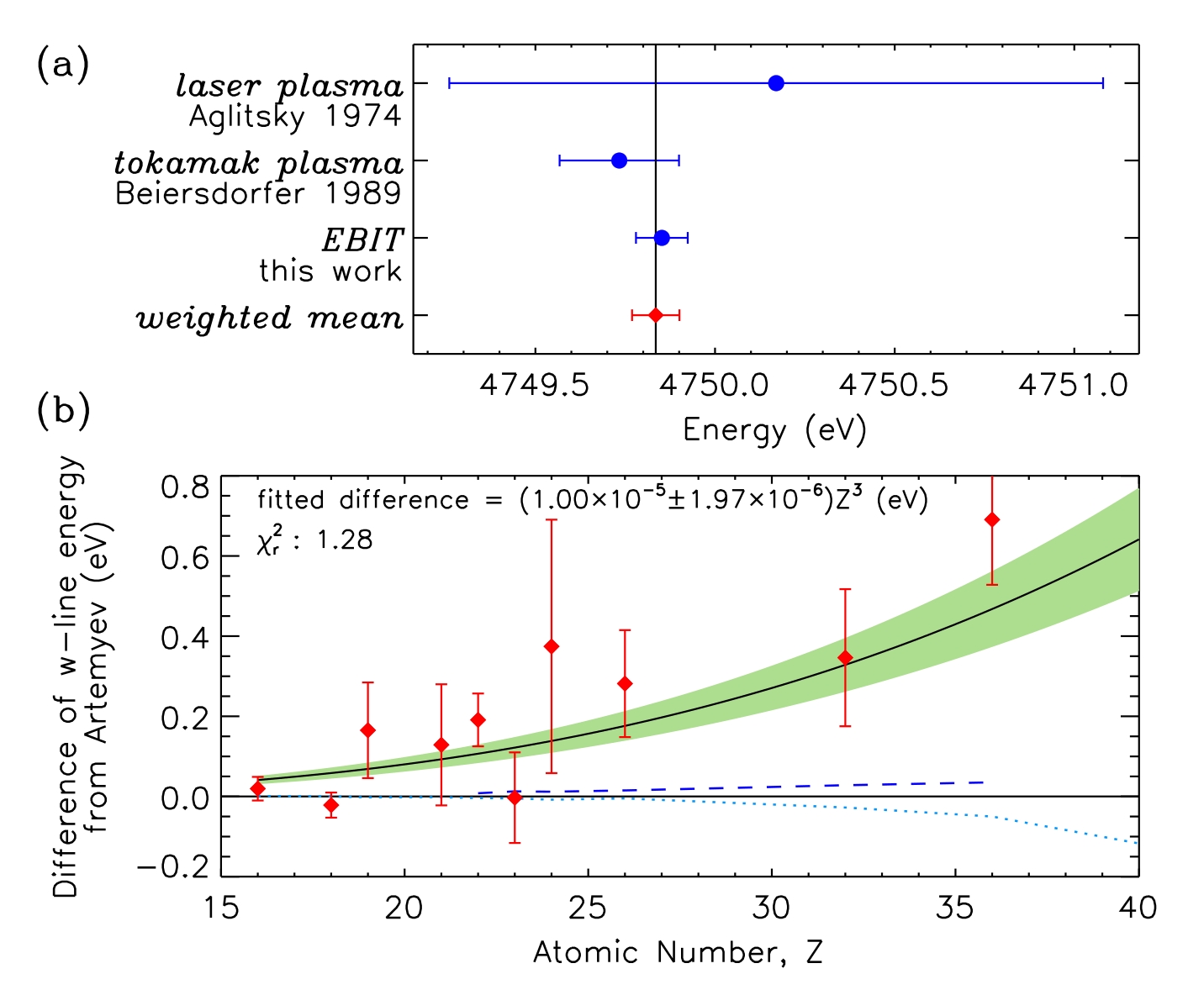}}
\end{picture}
\renewcommand{\normalsize}{\footnotesize}
    \caption{(a) The experimental data for He-like Ti. (b) The difference between experimental and theoretical K$_\alpha$ X lines for medium-heavy helium like ions). (From Chantler \it{et al.}~\cite{Chantler12},  in eV.)}
  \renewcommand{\normalsize}{\footnotesize}
 \label {Fig:Chantler12}
\end{figure}

Furthermore, Chantler \it{et al.}~\cite{Chantler12} have recently found "\it{Evidence for a Z-dependent Divergence Between Experiment and Calculation}" for the $K_\alpha$ lines in medium-heavy ions of several times the combined experimental and theoretical uncertainties (see Fig. \ref{Fig:Chantler12}).  The result for He-like Ti, studied by Chantler \it{et al.} at the EBIT facility at NIST is shown in Table \ref{Tab:Ti}. Here, the difference between theory and experiment is three times the experimental uncertainty. 
The theoretical results are obtained by Artemyev \it{et al.}~\cite{AShab05} and by Plante \it{et al.}~\cite{PJS94}, and the former claim much higher accuracy than the experimental result. 

The calculations of Artemyev \it{et al.} include two-photon exchange in a comprehensive way but do not contain electron correlation beyond second order. The calculations of Plante \it{et al.} contain all-order electron correlation effects but are less complete in the QED part. 

It is evident that it would be possible to test QED effects on a deeper level, were more accurate theoretical results available. We have recently developed a procedure for energy-dependent perturbation theory~\cite{ILBook11,LSH11,DanielTh10,HLS09,LSH08,HSL07,LSH07B}, which for the first time will make it possible to include QED perturbations in an all-order perturbation expansion in a rigorous way. This will take account of new effects beyond two-photon exchange that have previously not been considered. This is expected to improve the theoretical accuracy considerably. 

Calculations of some higher-order QED-MBPT effects of the type mentioned above have been performed at our laboratory, so far restricted to the ground state of He-like ions. In his Ph.D. thesis work Daniel Hedendahl~\cite{DanielTh10} calculated for the first time so-called non-radiative QED interactions (retardation and virtual pairs, see Fig. \ref{Fig:QEDEff}), combined with all-order electron correlation (see Table \ref{Tab:QEDCorr} below). The inclusion of radiative QED effects (self-energy, vertex correction, and vacuum polarization) into the all-order perturbative expansion is presently under way at our laboratory.

\begin{table}\normalsize
\begin{center}
\caption{\normalsize The transition $1s2s\,^1S_0 - 1s2p\,^3P_1$
for He-like Si (in cm$^{-1}$, 1eV=8066 cm$^{-1}$)}
\begin{tabular}{|l|l|l||}
\hline   & & Reference \\
\hline
Expt'l & 7230.585(6)& Myers \it{et al.}~\cite{Myers08} \\
Theory & 7229(2)&Artemyev \it{et al.}~\cite{AShab05}\\
 & 7231.1 & Plante \it{et al.}~\cite{PJS94}\\
 \hline
\end{tabular}
\label{Tab:Si} \end{center}
\end{table}

\begin{table}\normalsize
\begin{center}
\caption{\normalsize K$\alpha$ X-ray data for copper  (in eV)}
\begin{tabular}{|l|l|l|l||}
\hline   &K$\alpha_1$ &K$\alpha_2$ &Reference \\
\hline
Expt'l & 8047.8237(26)& 8027.8416(26)&Deslattes \it{et al.}~\cite{Deslattes03} \\
Theory  & 8047.86(4) &8027.92(4)&Chantler \it{et al.}~\cite{Chantler10}\\
\hline
\end{tabular}\label{Tab:CuXray}
\end{center}
\end{table}

\begin{table}\normalsize
\begin{center}
\caption{\normalsize The transition $1s^2\,^1S_0 - 1s2p\,^1P_1$
for He-like Ti  (in eV)}
\begin{tabular}{|l|l|l||}
\hline   & & Reference \\
\hline
Expt'l & 4749.85(7)& Chantler \it{et al.}~\cite{Chantler12}\\
Theory& 4749.644(1)&Artemyev \it{et al.}~\cite{AShab05}\\  & 4749.639 & Plante \it{et al.}~\cite{PJS94}\\
\hline
\end{tabular}
\label{Tab:Ti} \end{center}
\end{table}

The outline of the paper is as follows. In the next section we shall summarize the well-known many-body perturbation theory for \itul{energy-independent} perturbations, which forms the basis for the project. In the following section we shall briefly describe the newly developed theory for \itul{energy-dependent} perturbations, which is the instrument for incorporating the QED effects into the all-order perturbation expansion in a rigorous manner. In the final section we shall indicate how this can be accomplished and give some numerical results.

\section{Standard MBPT}\label{Sec:MBPT}
\subsection{Non-relativistic MBPT}
As a background we begin by summarizing the standard time-independent perturbation theory (see, for instance~\cite{LM86}). We consider a number of solutions to the non-relativistic Schr\šdinger equation, known as "target states",
\begin{equation}\label{Part}
  H\Psi^\alpha=E^\alpha\Psi^\alpha,
\end{equation}
where the Hamiltonian is given by
\begin{equation}\label{Ham}
  H=\sum_{i=1}^N h_S(i)+\VC.
\end{equation}
Here, $h_S$ is the Schr\šdinger single-electron Hamiltonian and $\VC$ is the electostatic interaction between the electrons
\begin{equation}\label{hS}
  \VC=\sum_{i<j}\frac{e^2}{4\pi\epsi_0\,r_{ij}}.
\end{equation}

For each target state there exists a zeroth-order or \it{model function}, $\Psi_0^\alpha$.
A wave operator, $\Omega$, transforms the latter to the exact solutions
\begin{equation}\label{WO}
  \Psi^\alpha=\Omega\Psi_0^\alpha.
\end{equation}
The model functions form a \it{model space}, and in \it{intermediate normalization} these functions are the projections of the exact solutions on this space
\begin{equation}\label{Proj}
  \Psi_0^\alpha=P\Psi^\alpha.
\end{equation}

We partition the Hamiltonian in the standard way into a model Hamiltonian and a perturbation 
\begin{equation}\label{Part2}
  H=H_0+V.
\end{equation}

The wave operator satisfies the \it{generalized Bloch equation}
\begin{equation}\label{Bloch}
  \boxed{\big[\Om,H_0\big]P=Q\big(V\Om-\Om W\big)P,}
\end{equation}
where $W$ is the \it{effective interaction}
\begin{equation}\label{Veff}
  W=V\eff=PV\Om P.
\end{equation}
This form of the Bloch equation is valid also when the model space contains different energy levels.

In the general case we can separate the Bloch equation into one equation for each energy level $\calE$ of the model space
\begin{equation}\label{Bloch2}
  \Om \PE=\GQ(\calE)\big(V\Om-\Om W\big)\PE,
\end{equation}
where $\GQ(\calE)=Q\Ga(\calE)$, $Q$ is the projection operator for the space outside the model space and
\begin{equation}\label{Ga}
    \Ga(\calE)=\sum_i\frac{\ket{i}\bra{i}}{\calE-E^0_i}
\end{equation}
is the \it{resolvent} with summation over all eigenstates $\ket{i} $ of $H_0$ with eigenvalue
$E^0_i$.  Expanding the equation \eqref{Bloch2} order by order, yields
\begin{equation}\label{Omn}
   \boxed{\Omega^{(n)}=  \GQ V \Omega^{(n-1)}-\GQ\sum_{m=1}^{n-1} \Omega^{(m)}W^{(n-m)},}
\end{equation}

The last term in the Bloch equation \eqref{Bloch2} and in the expansion \eqref{Omn} is usually referred to as "folded", since it is conventionally represented by folded diagrams. It can also
be regarded as the finite remainder, when the singularities---due to intermediate model-space states---of a ladder expansion are eliminated, which we shall illustrate as follows.

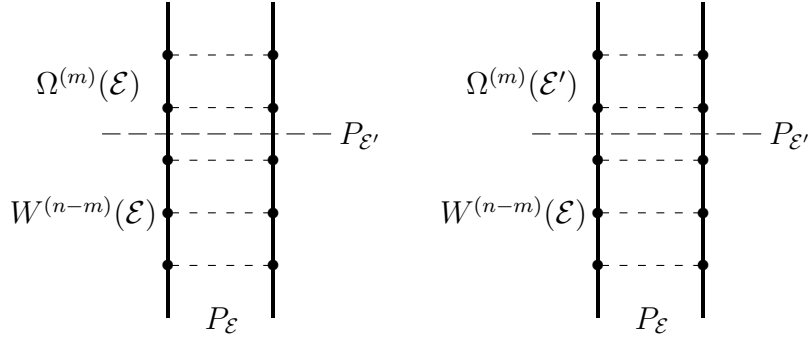
\begin{figure}
\begin{center}
\begin{picture}(8,3.5)(-1,0)
 \put(0,0){\LineV{6}} \put(2,0){\LineV{6}}
 \put(-2.5,4.25){\large$\Om^{(m)}(\calE)$}
  \put(-3,1.8){\large$W^{(n-m)}(\calE)$}
  \put(3.25,3.3){\large$\PEP$}\put(0.7,-0.2){\large$\PE$}
 \put(-1.6,3.5){\multiput(0.35,0)(0.5,0){9}{\line(1,0){0.35}}}
 \put(0,4){\elstat{}{}{}}\put(0,5){\elstat{}{}{}}
 \put(0,1){\elstat{}{}{}}\put(0,2){\elstat{}{}{}} \put(0,3){\elstat{}{}{}}
\end{picture}
\begin{picture}(4,3.5)(-1,0)
 \put(0,0){\LineV{6}} \put(2,0){\LineV{6}}
  \put(-2.5,4.25){\large$\Om^{(m)}(\calE')$}
   \put(-3,1.8){\large$W^{(n-m)}(\calE)$}
    \put(3.25,3.3){\large$\PEP$}\put(0.7,-0.2){\large$\PE$}
 \put(-1.6,3.5){\multiput(0.35,0)(0.5,0){9}{\line(1,0){0.35}}}
 \put(0,4){\elstat{}{}{}}\put(0,5){\elstat{}{}{}}
 \put(0,1){\elstat{}{}{}}\put(0,2){\elstat{}{}{}} \put(0,3){\elstat{}{}{}}
\end{picture}
 \renewcommand{\normalsize}{\footnotesize}
    \caption{Two-electron ladder diagram (left). In the corresponding counterterm (right) the energy parameter of the wave-operator part is changed to the energy of the intermediate model-space state ($\calE'$). The vertical lines represent electron orbitals/propagators and the dached lines the electrostatic interaction. Time is supposed to flow upwards.}
   \renewcommand{\normalsize}{\standard}
    \label{Fig:Ladder}
\end{center}
\end{figure}

We assume we have a wave-operator in the form of a "ladder" expansion (see Fig. \ref{Fig:Ladder}, left)
\begin{equation}\label{Ladd3}
    \Ga(\calE)V \Ga(\calE)V \Ga\calE)V\cdots \PE,
\end{equation}
where one of the intermediate states lies in the model space with an energy $\calE'$ ($\PEP$), almost degenerate with the state we start from with energy $\calE$. The ladder then becomes
\begin{equation}\label{LaddQ} 
    \GQ(\calE)V\cdots\frac{\PEP}{\calE-\calE'}V\GQ(\calE)V\cdots\PE,
\end{equation}
which can be expressed
\begin{equation}\label{LaddQ2}
    \Om^{(m)}(\calE)\frac{1}{\calE-\calE'}W^{(n-m)}(\calE)\PE,
\end{equation}
assuming there are $n$ interactions in total and  $m$ after the degeneracy. Since we have assumed that the interactions are energy independent, the wave operator and effective interaction depend on the energy only through the resolvents.

The expressions above are quasi-singular, and to eliminate the singularity we add a \it{counterterm},  (see Fig. \ref{Fig:Ladder}, right)
\begin{equation}\label{Counter}
    -\Om^{(m)}(\calE')\frac{1}{\calE-\calE'}W^{(n-m)}(\calE)\PE,
\end{equation}
where the energy of the wave operator is modified to $\calE'$. But the first term in the Bloch expansion \eqref{Omn} contains counterterms from previous orders, namely
\begin{equation}\label{Counter1}
  - \GQ(\calE)V\Om^{(m-1)}(\calE')\frac{1}{\calE-\calE'}\,W^{(n-m)}(\calE)\PE,
\end{equation}
The difference then becomes
\begin{equation}\label{Counter2}
  \Big(\GQ(\calE)-\GQ(\calE')\Big)V\Om^{(m-1)}(\calE')\frac{1}{\calE-\calE'}\,W^{(n-m)}(\calE)\PE
\end{equation}
or
\begin{equation}\label{}
 -\GQ(\calE)\Om^{(m)}(\calE)\,W^{(n-m)}(\calE)\PE,
\end{equation}
which is the folded term in the expansion \eqref{Omn}.

If we assume that the interactions can be different in each order of the expansion, $V_1,V_2,V_3,...$, then it follows that $\Omega^{(m)}$ should be formed by the $m$ \it{last interactions} and $W^{(n-m)}$ by the remaining ones.

\subsection{All-order expansion}
By separating the wave operator into one-, two-,... body parts,
\begin{equation}\label{OmSep}
   \Om=1+\Om_1+\Om_2+\cdots,
\end{equation}
the Bloch equation \eqref{Bloch} can be separated into a number of coupled equations,
\begin{equation}\label{Blochn}
  \big[\Om_n,H_0\big]P=Q\big(V\Om-\Om W\big)_nP.
\end{equation}
Normally, the two-body part dominates heavily, and this part leads to the \it{pair equation}.
Efficient numerical methods have been developed for solving this equation for atomic systems~\cite{SO89ab,SJ96}. Solving this equation iteratively (neglecting the minor one-body part), leads to the \it{all-order pair function}, illustrated in Fig. \ref{Fig:IPair}. For a two-electron system, starting from hydrogenic orbitals with no electron core, this will ultimately lead to the \it{exact non-relativistic wave function}~\cite{Ma79}.

\subsection{Relativistic MBPT. QED effects.}
\begin{figure}
\begin{center}
\setlength{\unitlength}{0.6cm}\hspace{0.8cm}
\begin{picture}(3.5,3)(1,0)
\small \put(0,1.5){\Elline{2.5}{1.75}{}{}}
\put(2,2.5){\Elline{1.5}{0.75}{}{}} \put(0,1.5){{\photonENE{
}{}{}}} \put(0,0){\Elline{1.5}{0.75}{}{}}
\put(2,0){\Elline{2.5}{0.75}{}{}}
\end{picture}
\begin{picture}(2,5)(1,0)
 \put(0,0){\Elline{1}{0.5}{}{}}\put(0,1){\Elline{2}{1}{}{}}
 \put(0,3){\Elline{1}{0.5}{}{}}
 \put(0,2){{\ElSE{}{}{}}}
\end{picture}
\begin{picture}(3,3)(0,0)
\tiny\put(0,1){\Elline{0.75}{0.25}{}{}}\put(0,1.75){\Elline{1.27}{0.75}{}{}}
\put(0,2.5){\Elline{1.5}{1}{}{}} \put(2,2.5){\Elline{1.5}{1}{}{}}
\put(0,2){{\ElSEL{}{}{}}}\put(0,1.75){{\photonENE{ }{}{}}}
\put(0,0){\Elline{1}{0.5}{}{}}\put(2,0){\Elline{2.5}{0.75}{}{}}
\end{picture}
\begin{picture}(2,3)(-2,0)
\small \put(0,0){\Elline{2}{1}{}{}}\put(0,2){\Elline{2}{1}{}{}}
\put(-1.5,2){{\VPloopL{}{}}} \put(-1,2){{\photonh}}
\end{picture}
\begin{picture}(3,6)(-1,0)
\tiny \put(0,0){\Elline{2}{1}{}{}}\put(0,2){\Elline{2}{1}{}{}}
\put(3,0){\Elline{2}{1}{}{}}\put(3,2){\Elline{2}{0.75}{}{}}
\put(0,2){{\photonSEG{}{}{}{}{}}}
\end{picture}
\begin{picture}(4,3)(-2,0)
\tiny\put(0,1){\Elline{2}{1}{}{}} \put(0,2.5){\Elline{1.5}{1}{}{}}
\put(0,2){\setlength{\unitlength}{0.4cm}{\SEpol{}{}{}{}{}}}
\put(0,0){\Elline{1}{0.5}{}{}}
\end{picture}
 \renewcommand{\normalsize}{\footnotesize}
    \caption{Effects beyond the NVPA are conventionally referred to as "QED effects". The first diagram represents retardation, which together with the effect of virtual electron-positron pairs is referred to as non-radiative QED effects. The following diagrams represent radiative effects. The second diagram represents the electron self-energy, the third vertex correction, and the last three vacuum polarization/photon self-energy. The internal  vertical lines represent electron propagators, with positive- and negative-energy orbitals.}
   \renewcommand{\normalsize}{\standard}
    \label{Fig:QEDEff}
    \end{center}
\end{figure}
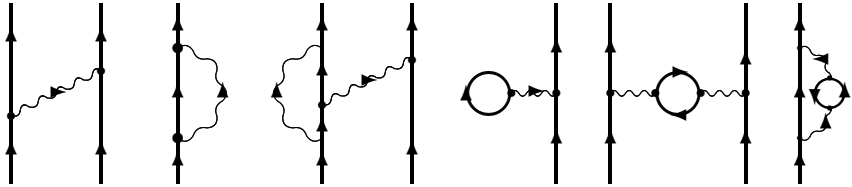

Relativistic many-body calculations normally start from the \it{projected
Dirac-Coulomb-Breit Hamiltonian}~\cite{Su80}, using the Coulomb gauge,
\begin{equation}\label{NVPA}
  H_\rm{NVPA}={\Lambda_+}{\Big[\sum_{i=1}^N h_D(i)+\VC
  +V_\rm{B}\Big]}{\Lambda_+},
\end{equation}
where
\begin{equation}\label{Breit}
  {V_\rm{B}=-\frac{e^2}{8\pi\epsi_0}\sum_{i<j}\Big[\frac{\balpha_i\bsdot\balpha_j}{\br_{ij}}
  + \frac{(\balpha_i\bsdot \br_{ij})(\balpha_j\bsdot \br_{ij})}{r_{ij}^2}\Big]}
\end{equation}
is the Breit interaction. $\balpha_i$ is the Dirac alpha matrix vector for particle i and
$\Lambda_+$ is a projection operator that excludes negative-energy states.
This is known as the \ul{\it{No-Virtual-Pair Approximation}} (NVPA). The all-order procedure described above can also be applied in this scheme.
Effects beyond NVPA are conventionally referred to as \it{QED effects} (see Fig. \ref{Fig:QEDEff}).

\begin{figure}
\begin{center}\setlength{\unitlength}{0.5cm}
\begin{picture}(4,4.5)(2,0)
 \put(0,0){\LineV{4}} \put(2,0){\LineV{4}}
  \put(0,2){\Pair{2}{}{}}
  \put(2.8,1.8){\Large =}
\end{picture}
 \begin{picture}(3.5,3.5)(2,0)
  \put(0,0){\LineV{4}} \put(2,0){\LineV{4}}
\put(2.5,1.8){\Large +}
\end{picture}
 \begin{picture}(3.5,3.5)(2,0)
  \put(0,0){\LineV{4}} \put(2,0){\LineV{4}}
    \put(0,2){\elstat{}{}{}}
\put(2.5,1.8){\Large +}
\end{picture}
\begin{picture}(3.5,3.5)(2,0)
  \put(0,0){\LineV{4}} \put(2,0){\LineV{4}}
    \put(0,1.33){\elstat{}{}{}}\put(0,2.67){\elstat{}{}{}} 
   \put(2.5,1.8){\Large +}
\end{picture}
\begin{picture}(4,3.5)(2,0)
 \put(0,0){\LineV{4}} \put(2,0){\LineV{4}}
    \put(0,1){\elstat{}{}{}}\put(0,2){\elstat{}{}{}} \put(0,3){\elstat{}{}{}}
   \put(2.5,1.8){\Large + $\cdots$}\put(5.2,1.8){{\Large + folded}}
\end{picture}
 \renewcommand{\normalsize}{\footnotesize}
    \caption{Graphical representation of the all-order Coulomb pair function. The "folded" term refers to the intermediate model-space contributions, discussed in the text. For a two-electron system with no electron core this represents exactly the non-relativistic correlation effect.}
   \renewcommand{\normalsize}{\standard}
    \label{Fig:IPair}
\end{center}
\end{figure}
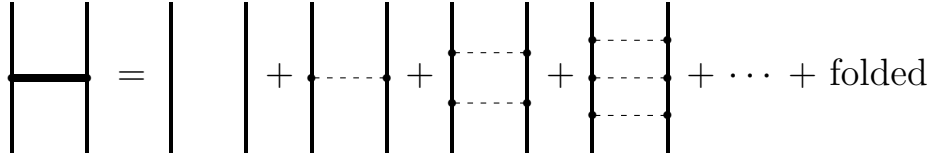

\section{Energy-dependent perturbation theory}
Next, we turn to time-dependent perturbation expansion in the form we have developed~\cite{ILBook11,LSH11,DanielTh10,HLS09,LSH08,HSL07,LSH07B}.
The standard \it{ time-evolution operator} describes the time evolution of the Schr\šdinger wave function in the \it{interaction picture} (IP)
\begin{equation}\label{EvolOp}
    \Psi_\mI(t)=U_\mI(t,t_0)\Psi_\mI(t_0).
\end{equation}
The IP is defined by the transformtion
\begin{equation}\label{IP}
    \Psi_\mI(t)=\me^{\im H_0t/\hbar}\Psi(t),
\end{equation}
where $\Psi$ represents the standard Schr\šdinger representation. The evolution operator satisfies the equation
\begin{equation}\label{Ueq}
    \im\hbar\Partder{t}{U_\mI(t,t_0)} =V_\mI U_\mI(t,t_0),
\end{equation}
where $V_\mI$ is the perturbation in IP.

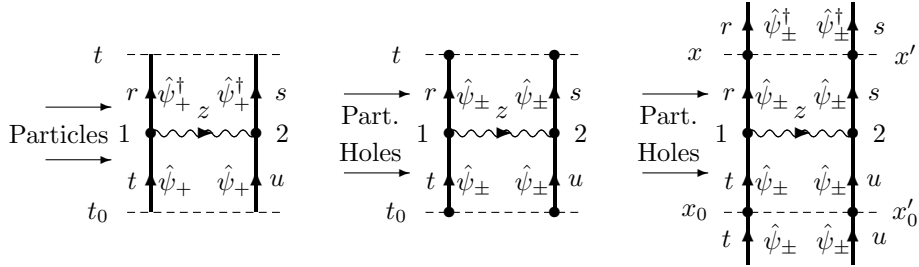
\begin{figure}
\begin{center}\setlength{\unitlength}{0.7cm}
\begin{picture}(5.5,)(-1,-1)
\put(-0.5,3){\dash{12}} \put(-1,3){\makebox(0,0){$t$}}
\put(0,0){\Elline{1.5}{0.6}{t}{\;\hpsi_+}}
\put(2,0){\Elline{1.5}{0.6}{\hpsi_+\;}{u}}
\put(0,1.5){\photon{z}{1}{2}}
 \put(0,1.5){\Elline{1.5}{0.75}{r}{\:\hpsi_+\dagg}}
 \put(2,1.5){\Elline{1.5}{0.75}{\hpsi_+\dagg\;}{\;s}}
 \put(-0.5,0){\dash{12}} \put(-1,0){\makebox(0,0){$t_0$}}
 \put(-2,2){\vector(1,0){1.25}}\put(-2,1){\vector(1,0){1.25}}
 \put(-1.75,1.5){\makebox(0,0){Particles}}
\end{picture}
\begin{picture}(5.5,5)(-1,-1)
\put(0,3){\dcirc{t}{}}
\put(0,0){\Elline{1.5}{0.6}{t}{\;\hpsi_\pm}}
\put(2,0){\Elline{1.5}{0.6}{\hpsi_\pm\;}{u}}
\put(0,1.5){\photon{z}{1}{2}}
 \put(0,1.5){\Elline{1.5}{0.75}{r}{\;\hpsi_\pm}}
 \put(2,1.5){\Elline{1.5}{0.75}{\hpsi_\pm\;}{s}}
\put(0,0){\dcirc{t_0}{}}
 \put(-2,2.25){\vector(1,0){1.25}}\put(-2,0.75){\vector(1,0){1.25}}
 \put(-1.5,1.85){\makebox(0,0){Part.}}\put(-1.5,1.15){\makebox(0,0){Holes}}
\end{picture}
\begin{picture}(3,5)(-1,-1)
\put(0,3){\dcircOut{x}{x'}{1}{r}{s}}
\put(0.3,3.4){$\hpsi_\pm\dagg$}\put(1.3,3.4){$\hpsi_\pm\dagg$}
\put(0,0){\Elline{1.5}{0.6}{t}{\;\hpsi_\pm}}
\put(2,0){\Elline{1.5}{0.6}{\hpsi_\pm\;}{u}}
\put(0,1.5){\photon{z}{1}{2}}
\put(0,1.5){\Elline{1.5}{0.75}{r}{\;\hpsi_\pm}}
 \put(2,1.5){\Elline{1.5}{0.75}{\hpsi_\pm\;}{s}}
\put(0,0){\dcircIn{x_0}{x'_0}{1}{t}{u}}
\put(0.3,-0.7){$\hpsi_\pm$}\put(1.3,-0.7){$\hpsi_\pm$}
 \put(-2,2.25){\vector(1,0){1.25}}\put(-2,0.75){\vector(1,0){1.25}}
 \put(-1.5,1.85){\makebox(0,0){Part.}}\put(-1.5,1.15){\makebox(0,0){Holes}}
\end{picture}
\renewcommand{\normalsize}{\footnotesize}
    \caption{Comparison between the standard evolution operator,
    the Green's function and the covariant evolution operator for
    single-photon exchange in the equal-time approximation. (From~\cite[Fig. 6.1]{ILBook11}).}
    \label{Fig:Comp}
\end{center}
\end{figure}

The standard evolution operator is relativistically \it{non-covariant}, since the time can flow only in the positive direction. It is for a two-particle system with a single photon exchange represented by the first diagram in Fig. \ref{Fig:Comp}. 

The single-particle Green's function can for a one-dimensional model space be defined
\begin{equation}\label{GF}
    G(t,t_0)=\frac{\bigbra{0_\mH}T[\hpsi_\mH(x)\hpsi_\mH\dagg(x_0)]\bigket{0_\mH}}
    {\bigbra{0_\mH}0_\mH\big\rangle},
\end{equation}
where $T$ is the Wick time-ordering operator and $\ket{0_\mH}$ is the vacuum state, and $\hpsi_\mH(x),\;\hpsi_\mH\dagg(x)$ are the electron-field operators in the Heisenberg representation (HP), defined by
\begin{equation}\label{HP}
    \Psi_\mH(t)=\me^{\im Ht/\hbar}\Psi(t).
\end{equation}

The numerator of \eqqref{GF} is normally singular, but the singularities are in this case eliminated by the denominator, so the ratio is regular.

The Green's function is relativistically \itul{covariant}. It is for a single-photon exchange between two electrons represented by the second diagram in Fig. \ref{Fig:Comp}. The free lines are represented by electron propagators, where time can flow in both directions.

We can define a \it{covariant evolution operator} (CEO) by generalizing the standard evolution operator, as illustrated for single-photon exchange in the last diagram in Fig. \ref{Fig:Comp}. Here, electron-operator lines are inserted at the free ends of the Green's-function diagram, transforming this into an operator diagram. 
The single-particle CEO (in IP) can be generally defined in analogy with the Green's function \eqqref{GF} as
\begin{eqnarray}\label{UCovDef}
   \U_\Cov(t,t_0)=\dint\dif^3\bx\,\dif^3\bx_0\,\hpsi_\mH\dagg(x)
   \bigbra{0_\mH}T[\hpsi_\mH(x)\hpsi_\mH\dagg(x_0)]\bigket{0_\mH}\hpsi_\mH(x_0).
\end{eqnarray}
This can be singular, like the numerator of the Green's-functions definition \eqref{GF}. Since we want the procedure here to be valid also for a general multi-dimensional model space, we shall use a different procedure for the regularization.
 
We define a \it{Green's operator} (GO), $\calG(t,t_0)$, by the relation (in the following omitting the subscript '$\Cov'$)
\begin{equation}\label{GO}
    U(t,t_0)P=\calG(t,t_0)\bdot PU(0,t_0)P.
\end{equation}
Here, the heavy dot implies that the denominators to the left of the dot are related to the model space at the position of the dot. This is analogous to the procedure we used in defining the counterterm in the time-independent case \eqref{Counter}. It can be shown that the GO is  \it{regular at all times}.

We assume that we have an \it{adiabatic damping}, so that at the limit $t\rarr-\infty$ the wave fundtion becomes an eigenfunction of the model Hamiltonian \eqref{Part},
\begin{equation}\label{Parent}
    \lim_{t\rarr-\infty}\Psi^\alpha(t)=\Phi^\alpha,
\end{equation}
a state we refer to as the \it{parent state}. According to the \it{Gell-Mann--Low theorem} (GML)~\cite{GML51,LSA04}, the wave function at $t=0$ can be expressed as the limit 
\begin{equation}\label{GML}
    \Psi^\alpha=\lim\frac{U(0,-\infty)\ket{\Phi^\alpha}}{\bra{\Psi_0^\alpha}U(0,-\infty)\ket{\Phi^\alpha}}.
\end{equation}
when the damping vanishes. Here, the numerator and the denominator are both normally singular, and only the ratio is regular.

We assume, in analogy with the non-relativistic case, that the CEO describes the evolution of the relativistic wave function
\begin{equation}\label{RelEvol}
    \Psi(t)=U(t,t_0)\Psi(t_0),
\end{equation}
so according to the Gell-Mann--Low theorem 
\begin{equation}\label{Psit}
    \Psi(t)=N_\alpha U(t,-\infty)\Phi_\alpha,
\end{equation}
where $N_\alpha$ is the normalization constant
\begin{equation}\label{Norm}
   N_\alpha=\frac{1}{\bra{\Psi_0^\alpha}U(0,-\infty)\ket{\Phi^\alpha}}.
\end{equation}

According to the definition \eqref{GO} we have
\begin{equation}\label{GO2}
    U(0,-\infty)P=\calG(0,-\infty)\bdot PU(0,-\infty)P,
\end{equation}
so the wave function at $t=0$ becomes
\begin{equation}
    \Psi^\alpha(0)=N_\alpha\calG(0,-\infty)\bdot PU(0,-\infty)\Phi_\alpha.
\end{equation}
But $N_\alpha PU(0,-\infty)\Phi_\alpha$ is according to the GML theorem equal to the model function \eqref{Proj}, which leads to the important relation
\begin{equation}
    \Psi^\alpha(0)=\calG(0,-\infty)\Psi^\alpha_0.
\end{equation}
Comparing with the relation \eqref{WO}, we then find that $\calG(0,-\infty)$ is the \it{analogue of the MBPT wave operator}
\begin{equation}\label{ROm}
   \boxed{ \Om=\calG(0,-\infty).}
\end{equation}
 
We shall now look at the perturbation expansion of the Green's operator \eqref{GO}. For $t=0$ we have $U^{(0)}P=\calG^{(0)}P=P$, which gives from the definition \eqref{GO2}
\begin{equation}\label{GO1}
    \calG\ett P=QU\ett P,
\end{equation}
\begin{equation}\label{GO2a}
    \calG\tva P=QU\tva P-\calG\ett\bdot PU\ett P,
\end{equation}
\begin{equation}\label{GO3}
    \calG\tre P=QU\tre P-\calG\ett\bdot PU\tva P-\calG\tva\bdot PU\ett P,
\end{equation}
etc. The negative terms are \it{counterterms} that eliminate the singularities of the evolution operator in close analogy with the treatment of the previous section.
Using $QU\tva=\GQV(\GQ+\GP) VP$ we have
  \[\calG\tva P=\calG_0\tva+\Delta(\calG\ett) PU\ett P,\]
  where $\calG_0=1+\GQ V+\GQ V\GQ V+\cdots$ is the GO without intermediate model-space states, and   \[\Delta(\calG\ett) PU\ett P=\calG\ett(\calE)\PEP U\ett \PE-\calG\ett(\calE') \PEP U\ett \PE.\]
Similarly, we have 
   \[\calG\tre P=\calG\ett Q U\tva P+\Delta(\calG\ett)PU\tva P-\calG\tva\bdot PU\ett P\]  

We want to express $\calG\tre P$ as
  \(\calG\tre P=\GQ V\calG\tva P+X\) 
  in order to find an analogue of the MBPT Bloch equation \eqref{Omn}. This gives
  
\(X= \calG_0\tva PU\ett P-\calG_0\tva\bdot PU\ett P+\Delta(\calG\ett)\Big(PU\tva P -PU\ett P\bdot PU\ett P\Big)-\calG\ett\Delta(\calG\ett)PU\ett P\)

\(= \Delta(\calG_0\tva )  W_0\ett+\Delta(\calG\ett)W_0\tva+\Delta\Big(\Delta(\calG\ett)W_0\ett\Big)W_0\ett-\calG\ett\Delta(\calG\ett)W_0\ett\)

\(= \Delta^*(\calG\tva)W_0\ett+\Delta(\calG\ett)\Big(W_0\tva+\Delta(W_0\ett)W_0\ett\Big),\)

\noindent where the asterisk indicates that the difference is taken only with respect to the last interaction $\GQ V$. This leads to
\begin{equation}
    \calG\tva=\GQ V\calG\ett+\partdelta{\calG\ett}{\calE}W_0\ett
\end{equation}
\begin{equation}
    \calG\tre=\GQ V\calG\tva+\partdelta{\calG\ett}{\calE}\big(W_0\tva+W_1\tva\big)
    +\pda{\calG\tva}W_0\ett,
\end{equation}
  which is consistent with the general equation~\cite[Eq. (66)]{LSH11},~\cite[6.106]{ILBook11} for $t=0$
  \begin{equation}\label{GenBloch}
 \boxed{ \calG=1+\GQ V\calG+\pda{\calG}\,W.}
\end{equation}
This can be expressed order by order as  \begin{equation}\label{GenBlochit}
  \boxed{\calG^{(n)}=\GQ V\calG^{(n-1)}+\sum_m\pda{\calG}^{(m)}\,W^{(n-m)}.}
\end{equation}
This is the Bloch equation for general energy-dependent perturbation.
For an energy-independent perturbation the differentiation of the last interaction $\GQ V$ leads to a factor of ($-\GQ$), and \eqqref{Omn} is retrieved. If the perturbations can be different in different orders, $\calG^{(m)}$ in the sum should be formed by the 
$m$ last interactions and $W$ by the remaining ones, also in complete analogy with the MBPT case \eqref{Omn}.

\section{Combined QED-MBPT}
With the energy-dependent perturbation theory that we have developed, anddescribed in the previous section, we can mix (one-photon) energy-dependent QED perturbations with Coulomb interactions to arbitrary order~\cite{LSH06,LSH11}. 

We employ the Coulomb gauge in order to be able to utilize as much as possible of the development in standard MBPT. This has among other things the advantage that we can represent most of the transverse-photon exchange by means of the instantaneous Breit interaction \eqqref{Breit}.

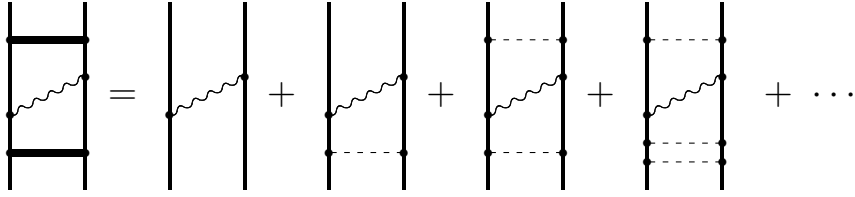
\begin{figure}
\begin{center}\setlength{\unitlength}{0.5cm}
\begin{picture}(4,4.5)(1,0)
 \put(0,0){\LineV{5}} \put(2,0){\LineV{5}}
  \put(0,1){\Pair{2}{}{}} \put(0,4){\Pair{2}{}{}}
  \put(0,2){\photonENEG{}{}{}}
  \put(2.6,2.3){\Large =}
\end{picture}
\begin{picture}(4,4.5)(1,0)
 \put(0,0){\LineV{5}} \put(2,0){\LineV{5}}
  \put(0,2){\photonENEG{}{}{}}
  \put(2.6,2.3){\Large +}
\end{picture}
\begin{picture}(4,4.5)(1,0)
 \put(0,0){\LineV{5}} \put(2,0){\LineV{5}}
  \put(0,2){\photonENEG{}{}{}}\put(0,1){\elstat{}{}{}} 
  \put(2.6,2.3){\Large +}
\end{picture}
\begin{picture}(4,4.5)(1,0)
 \put(0,0){\LineV{5}} \put(2,0){\LineV{5}}
  \put(0,2){\photonENEG{}{}{}}\put(0,1){\elstat{}{}{}} \put(0,4){\elstat{}{}{}} 
  \put(2.6,2.3){\Large +}
\end{picture}
\begin{picture}(4,4.5)(1,0)
 \put(0,0){\LineV{5}} \put(2,0){\LineV{5}}
  \put(0,2){\photonENEG{}{}{}}\put(0,0.75){\elstat{}{}{}} \put(0,1.25){\elstat{}{}{}} \put(0,4){\elstat{}{}{}} 
  \put(3.1,2.3){\Large +\hhsp$\cdots$}
\end{picture}
 \renewcommand{\normalsize}{\footnotesize}
    \caption{Graphical representation of the all-order pair function combined with a transverse (Breit) interaction. The first two diagrams represent one- and two-photon exchange that have previously been calculated, while the remaining ones represent higher-order effects not calculated before.}
   \renewcommand{\normalsize}{\standard}
    \label{Fig:Breit}
\end{center}
\end{figure}

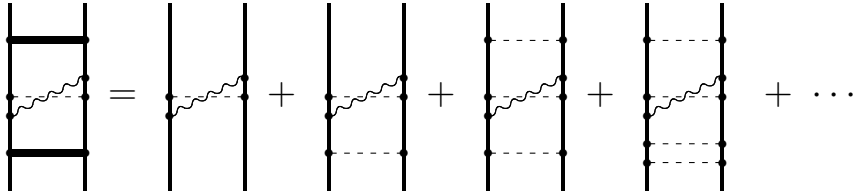
\begin{figure}
\begin{center}\setlength{\unitlength}{0.5cm}
\begin{picture}(4,4.5)(1,0)
 \put(0,0){\LineV{5}} \put(2,0){\LineV{5}}
  \put(0,1){\Pair{2}{}{}} \put(0,4){\Pair{2}{}{}}
  \put(0,2){\photonENEG{}{}{}}\put(0,2.5){\elstat{}{}{}} 
  \put(2.6,2.3){\Large =}
\end{picture}
\begin{picture}(4,4.5)(1,0)
 \put(0,0){\LineV{5}} \put(2,0){\LineV{5}}
  \put(0,2){\photonENEG{}{}{}}\put(0,2.5){\elstat{}{}{}}
  \put(2.6,2.3){\Large +}
\end{picture}
\begin{picture}(4,4.5)(1,0)
 \put(0,0){\LineV{5}} \put(2,0){\LineV{5}}
  \put(0,2){\photonENEG{}{}{}}\put(0,2.5){\elstat{}{}{}}\put(0,1){\elstat{}{}{}} 
  \put(2.6,2.3){\Large +}
\end{picture}
\begin{picture}(4,4.5)(1,0)
 \put(0,0){\LineV{5}} \put(2,0){\LineV{5}}
  \put(0,2){\photonENEG{}{}{}}\put(0,2.5){\elstat{}{}{}}\put(0,1){\elstat{}{}{}} \put(0,4){\elstat{}{}{}} 
  \put(2.6,2.3){\Large +}
\end{picture}
\begin{picture}(4,4.5)(1,0)
 \put(0,0){\LineV{5}} \put(2,0){\LineV{5}}
  \put(0,2){\photonENEG{}{}{}}\put(0,2.5){\elstat{}{}{}}\put(0,0.75){\elstat{}{}{}} \put(0,1.25){\elstat{}{}{}} \put(0,4){\elstat{}{}{}} 
  \put(3.1,2.3){\Large +\hhsp$\cdots$}
\end{picture}
 \renewcommand{\normalsize}{\footnotesize}
    \caption{Same as Fig. \ref{Fig:Breit} with a crossing  Coulomb interaction.}
   \renewcommand{\normalsize}{\standard}
    \label{Fig:BreitX}
\end{center}
\end{figure}

In Fig. \ref{Fig:Breit} we illustrate the incorporation of a transverse-photon exchange, i.e., Breit interaction (retarded and unretarded), into an all-order pair function, and in Fig. \ref{Fig:BreitX} the same when also a Coulomb interaction is crossing the Breit interaction. By replacing one of the Coulomb interactions by an instantaneous Breit interaction, we can also take account of the double Breit interaction where one interaction is instantaneous and one is retarded. Furthermore, multiple Coulomb crossings as well as virtual pairs (negative energy states) can be handled. These two sets represent the leading \it{non-radiative} QED interactions in combination with all-order electron correlation. They have been evaluated for He-like ions in the ground state, $1s^2$,  by Hedendahl in his PhD thesis~\cite{DanielTh10}. The diagrams beyond second order had never been calculated before. Some results are shown in Table \ref{Tab:QEDCorr}. 

\begin{table}\normalsize
\begin{center}
\caption{\normalsize QED-correlation effects for He-like ions beyond two-photon exchange (first four columns from Hedendahl~\cite{DanielTh10}, in eV).}
\begin{tabular}{||c|c|c|c|c|c||}
\hline   Z &Full Breit&Ret. part&Cross.Coul&Virt.Pairs &Self-energy \\
\hline
10 &0.0061& -0.0011&-0.0006&0.0002& \\
14 &0.0082& -0.0019&-0.0010&0.0004&\\
18 &0.0191 &-0.0027&-0.0014&0.0006&(-0.004)\\
24&0.0127&-0.0042&-0.0021&&\\
30&0.0150&-0.0057&-0.0028&0.0014&\\
36&&&&&(-0.010)\\
42&0.0187&-0.0087&-0.004&0.0019&\\
50&0.0215&-0.011&-0.005&0.0024&\\
66&0.025&-0.015&-0.006&0.0030&\\
\hline
\end{tabular}
\label{Tab:QEDCorr} \end{center}
\end{table}

\begin{figure}
\begin{center}\setlength{\unitlength}{0.5cm}
\begin{picture}(4,5)(1,0)
 \put(0,0){\LineV{5}} \put(2,0){\LineV{5}}
  \put(0,1){\Pair{2}{}{}} \put(0,4){\Pair{2}{}{}}
 \put(0,2.5){\setlength{\unitlength}{0.4cm}\ElSEG{}{}{}}
  \put(2.6,2.3){\Large =}
\end{picture}
\begin{picture}(4,4.5)(1,0)
 \put(0,0){\LineV{5}} \put(2,0){\LineV{5}}
  \put(0,2.5){\setlength{\unitlength}{0.4cm}\ElSEG{}{}{}}
  \put(2.6,2.3){\Large +}
\end{picture}
\begin{picture}(4,4.5)(1,0)
 \put(0,0){\LineV{5}} \put(2,0){\LineV{5}}
  \put(0,2.5){\setlength{\unitlength}{0.4cm}\ElSEG{}{}{}}
\put(0,1){\elstat{}{}{}} 
  \put(2.6,2.3){\Large +}
\end{picture}
\begin{picture}(4,4.5)(1,0)
 \put(0,0){\LineV{5}} \put(2,0){\LineV{5}}
 \put(0,2.5){\setlength{\unitlength}{0.4cm}\ElSEG{}{}{}}
\put(0,1){\elstat{}{}{}} \put(0,4){\elstat{}{}{}} 
  \put(2.6,2.3){\Large +}
\end{picture}
\begin{picture}(4,4.5)(1,0)
 \put(0,0){\LineV{5}} \put(2,0){\LineV{5}}
 \put(0,2.5){\setlength{\unitlength}{0.4cm}\ElSEG{}{}{}}
\put(0,0.75){\elstat{}{}{}} \put(0,1.25){\elstat{}{}{}} \put(0,4){\elstat{}{}{}} 
  \put(3.1,2.3){\Large +\hhsp$\cdots$}
\end{picture}
 \renewcommand{\normalsize}{\footnotesize}
    \caption{Graphical representation of the all-order pair function combined with the electron self-energy.}
   \renewcommand{\normalsize}{\standard}
    \label{Fig:SE}
\end{center}
\end{figure}
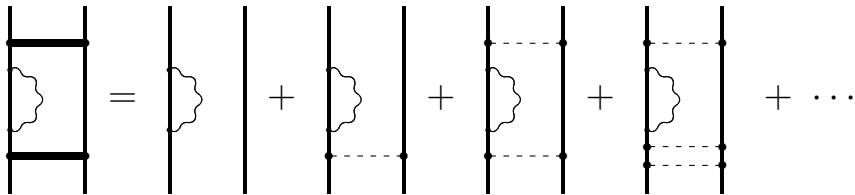

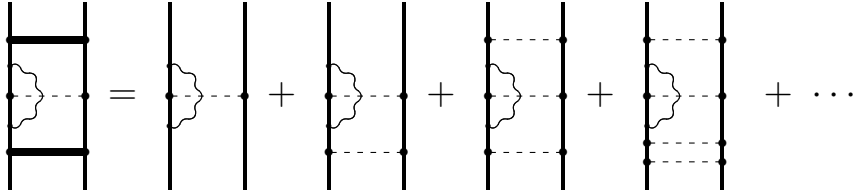
\begin{figure}
\begin{center}\setlength{\unitlength}{0.5cm}
\begin{picture}(4,4.5)(1,0)
 \put(0,0){\LineV{5}} \put(2,0){\LineV{5}}
  \put(0,1){\Pair{2}{}{}} \put(0,4){\Pair{2}{}{}}
 \put(0,2.5){\setlength{\unitlength}{0.4cm}\ElSEG{}{}{}}\put(0,2.5){\elstat{}{}{}} 
  \put(2.6,2.3){\Large =}
\end{picture}
\begin{picture}(4,4.5)(1,0)
 \put(0,0){\LineV{5}} \put(2,0){\LineV{5}}
  \put(0,2.5){\setlength{\unitlength}{0.4cm}\ElSEG{}{}{}}\put(0,2.5){\elstat{}{}{}} 
  \put(2.6,2.3){\Large +}
\end{picture}
\begin{picture}(4,4.5)(1,0)
 \put(0,0){\LineV{5}} \put(2,0){\LineV{5}}
  \put(0,2.5){\setlength{\unitlength}{0.4cm}\ElSEG{}{}{}}\put(0,2.5){\elstat{}{}{}} 
\put(0,1){\elstat{}{}{}} 
  \put(2.6,2.3){\Large +}
\end{picture}
\begin{picture}(4,4.5)(1,0)
 \put(0,0){\LineV{5}} \put(2,0){\LineV{5}}
 \put(0,2.5){\setlength{\unitlength}{0.4cm}\ElSEG{}{}{}}\put(0,2.5){\elstat{}{}{}} 
\put(0,1){\elstat{}{}{}} \put(0,4){\elstat{}{}{}} 
  \put(2.6,2.3){\Large +}
\end{picture}
\begin{picture}(4,4.5)(1,0)
 \put(0,0){\LineV{5}} \put(2,0){\LineV{5}}
 \put(0,2.5){\setlength{\unitlength}{0.4cm}\ElSEG{}{}{}}
\put(0,0.75){\elstat{}{}{}} \put(0,1.25){\elstat{}{}{}} \put(0,4){\elstat{}{}{}} \put(0,2.5){\elstat{}{}{}} 
  \put(3.1,2.3){\Large +\hhsp$\cdots$}
\end{picture}
 \renewcommand{\normalsize}{\footnotesize}
    \caption{Graphical representation of the all-order pair function combined with the Coulomb vertex correction.}
   \renewcommand{\normalsize}{\standard}
    \label{Fig:SEX}
\end{center}
\end{figure}

To insert the radiative effects into the all-order pair function is more difficult, but such work is now in progress at our laboratory, and some preliminary results for the self-energy part are inserted in Table \ref{Tab:QEDCorr}. As mentioned, it is advantageous to use the Coulomb gauge in this type of calculation, but numerical evaluation of the self-energy in this gauge has not been done until quite recently. The first calculation of this kind was performed in 2011 for  hydrogenlike ions by Hedendahl and Holmberg~\cite{DanJoh11}. Inserting the self-energy and the vertex correction into the Coulomb pair function,  leads to the sequences illustrated in Figs \ref{Fig:SE} and  \ref{Fig:SEX}. 

\begin{figure}
\begin{center}
\setlength{\unitlength}{0.5cm}
\begin{picture}(4.5,7)(-0.5,0.5) 
\put(0,1){\LineV{5}}\put(2,1){\LineV{5}}
\put(0,2){{\photonENEG{}{}{}}}
\put(0,4){{\photonENEG{}{}{}}}
 \end{picture}\begin{picture}(4.5,7)(-0.5,0) 
\put(0,0){\LineV{6}}\put(2,0){\LineV{6}}
\put(0,1.25){\Elstat{}{}{}}
\put(0,2){{\photonENEG{}{}{}}}
\put(0,4){{\photonENEG{}{}{}}}
 \end{picture}\begin{picture}(4.5,7)(-0.5,0) 
\put(0,0){\LineV{6}}\put(2,0){\LineV{6}}
\put(0,0.75){\Elstat{}{}{}}\put(0,1.25){\Elstat{}{}{}}\put(0,3.5){\Elstat{}{}{}}
\put(0,2){{\photonENEG{}{}{}}}
\put(0,4){{\photonENEG{}{}{}}}
 \end{picture}
\begin{picture}(4.5,2.5)(-0.5,0)
\put(0,0.75){\Elstat{}{}{}}\put(0,6){\Elstat{}{}{}}\put(0,3){\Elstat{}{}{}}
\put(0,4.5){\ElSELG{}{}{}}
\put(0,1.5){{\photonENEG{}{}{}}}
\put(0,0){\LineV{6.5}} \put(2,0){\LineV{6.5}}
\end{picture}
\begin{picture}(3,2.5)(-0.5,0)
\put(0,0.75){\Elstat{}{}{}}\put(0,6){\photonENEG{}{}{}}\put(0,3){\Elstat{}{}{}}
\put(0,4.5){\ElSELG{}{}{}}
\put(0,1.5){{\photonENEG{}{}{}}}
\put(0,0){\LineV{7.5}} \put(2,0){\LineV{7.5}}
\end{picture}
 \renewcommand{\normalsize}{\footnotesize}
    \caption{Examples of reducible multi-photon diagrams that can be evaluated by iterating the procedure with various QED perturbations together with Coulomb interactions. So far, only the first two-photon diagram has been calculated.}
  \renewcommand{\normalsize}{\footnotesize}
   \label{Fig:Red}
\end{center}
\end{figure}
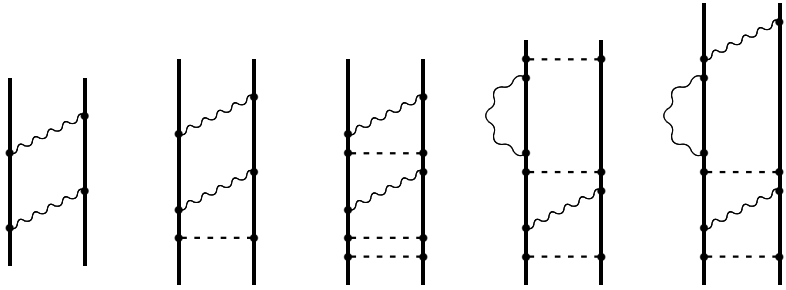

\begin{figure}
\begin{center}
\setlength{\unitlength}{0.5cm}
\begin{picture}(5,5)(-0.5,0) \small
\put(0,0.25){\Pair{2}{}{}}\put(0,4){\Pair{2}{}{}}
\put(0,0.75){{\photonNEG{}{}{}}}
\put(0,1.25){{\photonNEG{}{}{}}} 
\put(0,-0.5){\LineEtt{5.5}{}}\put(2,-0.5){\LineV{5.5}}
\end{picture}
\begin{picture}(5,3)(-0.5,0)
\put(0,0.25){\Pair{2}{}{}}\put(0,4){\Pair{2}{}{}}
\put(0,1){{\CrossphotonsG{ }{}{}{}{}{}}}
\put(0,-0.5){\LineV{5.5}}\put(2,-0.5){\LineV{5.5}}
\end{picture}
\begin{picture}(3,2.5)(-0.5,0)
\put(0,0.25){\Pair{2}{}{}}\put(0,4){\Pair{2}{}{}}
\put(0,2){{\ElSELG{}{}{}}}\put(0,1.75){{\photonENEG{}{}{}}}
\put(0,-0.5){\LineV{5.5}}\put(2,-0.5){\LineV{5.5}}
\end{picture}
 \renewcommand{\normalsize}{\footnotesize}
    \caption{Irreducible retarded multi-photon effects can for the time being not be included for computational reasons. Most of the effect can be included, though, by replacing one of the reatarded Breit interactions by the instantaneous one.}
  \renewcommand{\normalsize}{\footnotesize}
   \label{Fig:Irred}
\end{center}
\end{figure}
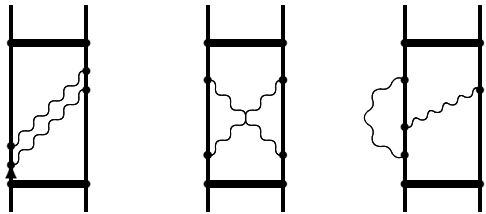

It is possible to insert several (single-photon) QED perturbations in the expansion, provided they are \it{reducible}, implying that a separated horizontal time line can be inserted between the individual interactions. A few examples of such combinations are given in Fig. \ref{Fig:Red}. Again, effects beyond the first two-photon diagram have never been accounted for in the past.

As mentioned, in using the Coulomb gauge, one or several of the Coulomb interactions can be replaced by instantaneous Breit interactions \eqqref{Breit}, and in this way most of the effect of the multi-photon exchange can be accounted for. Therefore, the procedure presented here can take care of all MBPT-QED effects, except \it{irreducible} ones, where more than one interaction is retarded, as illustrated in Fig. \ref{Fig:Irred}. These parts, however, which are inaccessible for the moment for computational reasons, are expected to be very small, representing, maybe, one or a few percent of the total two-photon effect for medium-heavy ions and much smaller than some of the higher-order effects in Figs \ref{Fig:Breit} to \ref{Fig:SEX} that the procedure does include. Furthermore, the pure two-photon part---without electron correlation---can here be evaluated by means of standard two-photon programs and added separately~\cite{PSS96,AShab05}. 

We have here considered only two-electron systems, but the procedure can be extended to many-electron systems in exactly the same way as in standard many-body theory. In addition, it can be combined with the effective \it{Coupled-Cluster Approach}, as recently demonstrated (CCA)~\cite{LSH10}.

\section{Conclusions}
An accurate procedure is now being developed for calculating energy levels and separations on few-electron highly charged ions. The procedure utilizes the newly developed energy-dependent perturbation theory and can thereby combine different effects of quantum-electrodynamics (QED) with electron correlation to all orders in a rigorous manner. The procedure will include important effects that have never been calculated before and is therefore expected to yield considerably higher accuracy than has previously been attainable. The procedure is currently being implemented and tested, to begin with on the ground-state of He-like ions. The purpose is to test the theory of QED for highly charged, few-electron ions on a higher level than has previously been possible. To what extent this might explain the discrepancies found by Chantler (Fig. \ref{Fig:Chantler12}) remains to be seen.

\section{Acknowledgements}
The authors are grateful to their former coworker Daniel Hedendahl for helpful discussions. This work has been supported in part by the Swedish Research Council, Vetenskapsr\aa det, and the Swedish National Infrastructure for Computing (SNIC).

\bibliographystyle{prsty}

\input{EPTTestQED.bbl}
\end{document}

%% file: Commands.tex
\newcommand{\setl[1]}{\setlength{\unitlength}{#1cm}}
\newcommand{\nn}{\nonumber \\}
\newcommand{\eqqref}[1]{\mbox{Eq. (\ref{#1})}}
\newcommand{\eqpref}[1]{\mbox{(Eq. \ref{#1})}}
\newcommand{\eps}{\ensuremath{\varepsilon}}
\newcommand{\me}{\mathrm{e}}
\newcommand{\im}{\ensuremath{\mathrm{i}}}
\newcommand{\bra}[1]{\langle #1 |}
\newcommand{\ket}[1]{| #1 \rangle}
\newcommand{\proj}[1]{\ket{#1}\bra{#1}}
\newcommand{\Bra}[1]{\left\langle #1 \left|}
\newcommand{\Ket}[1]{\right| #1 \right\rangle}
\newcommand{\bigbra}[1] {\big\langle #1\big|}
\newcommand{\bigket}[1] {\big|#1\big\rangle}
\newcommand{\Bigbra}[1] {\Big\langle #1\Big|}
\newcommand{\Biggbra}[1] {\Bigg\langle #1\Bigg|}
\newcommand{\Bigket}[1] {\Big|#1\Big\rangle}
\newcommand{\Biggket}[1] {\Bigg|#1\Bigg\rangle}
\newcommand{\dif}{\ensuremath{\mathrm{d}}}
\newcommand{\difbx}{\dif^3\bx}
\newcommand{\difbk}{\dif^3\bk}
\newcommand{\difbp}{\dif^3\bp}
\newcommand{\intDim}[1]{\int\frac{\dif^D #1}{(2\pi)^D}}
\newcommand{\intD}[1]{\int\frac{\dif^3 #1}{(2\pi)^3}}
\newcommand{\intDD}[1]{\int\frac{\dif^4 #1}{(2\pi)^4}}
\newcommand{\intdinf}[1]{\int_{-\infty}^\infty\frac{\dif #1}{2\pi}}
\newcommand{\intbx}{\int\dif^3\bx}
\newcommand{\bx}{\boldsymbol{x}}
\newcommand{\bxdot}{\dot{\bx}}
\newcommand{\br}{\boldsymbol{r}}
\newcommand{\bk}{\boldsymbol{k}}
\newcommand{\bp}{\bs{\rm{p}}}
\newcommand{\bpi}{\boldsymbol{\pi}}
\newcommand{\bq}{\bs{\rm{q}}}
\newcommand{\bl}{\bs{\rm{l}}}
\newcommand{\bj}{\bs{\rm{j}}}
\newcommand{\bsp}{\bs{\rm{s}}}
\newcommand{\bL}{\bs{\rm{L}}}
\newcommand{\bJ}{\bs{\rm{J}}}
\newcommand{\bI}{\bs{\rm{I}}}
\newcommand{\bF}{\bs{\rm{F}}}
\newcommand{\bS}{\bs{\rm{S}}}
\newcommand{\hatl}{\hat{l}}
\newcommand{\hats}{\hat{s}}
\newcommand{\hatj}{\hat{j}}
\newcommand{\hatbj}{\hat{\bs{\rm{j}}}}
\newcommand{\hatbl}{\hat{\bs{\rm{l}}}}
\newcommand{\hatbs}{\hat{\bs{\rm{s}}}}
\newcommand{\hatL}{\hat{L}}
\newcommand{\hatS}{\hat{S}}
\newcommand{\hatJ}{\hat{J}}
\newcommand{\hatbJ}{\hat{\bs{\rm{J}}}}
\newcommand{\hatbL}{\hat{\bs{\rm{L}}}}
\newcommand{\hatbS}{\hat{\bs{\rm{S}}}}
\newcommand{\hatH}{\hat{H}}
\newcommand{\hrho}{\hat{\rho}}
\newcommand{\balpha}{\ensuremath{\boldsymbol{\alpha}}}
\newcommand{\bgamma}{\bs{\rm{\gamma}}}
\newcommand{\bgd}{\bs{\rm{\gamma}}\bsdot\!}
\newcommand{\bgp}{\bs{\rm{\gamma}}\bsdot\bp}
\newcommand{\bgq}{\bs{\rm{\gamma}}\bsdot\bq}
\newcommand{\bgk}{\bs{\rm{\gamma}}\bsdot\bk}
\newcommand{\gnp}{\gamma^0p_0}
\newcommand{\gnk}{\gamma^0k_0}
\newcommand{\giki}{\gamma^ik_i}
\newcommand{\gipi}{\gamma^ip_i}
\newcommand{\gmu}{\gamma^\mu}
\newcommand{\gnu}{\gamma^\nu}
\newcommand{\gs}{\gamma^\sigma}
\newcommand{\ga}{\gamma}
\newcommand{\wts}{\wt{\gamma}^\sigma}
\newcommand{\wtg}{\wt{\gamma}}
\newcommand{\gt}{\gamma^\tau}
\newcommand{\gb}{\gamma^\beta}
\newcommand{\gn}{\gamma^0}
\newcommand{\gi}{\gamma^i}
\newcommand{\gii}{\gamma_i}
\newcommand{\gj}{\gamma^j}
\newcommand{\gjj}{\gamma_j}
\newcommand{\gE}{\gamma_\rm{E}}
\newcommand{\dd}[1]{\frac{\partial}{\partial #1}}
\newcommand{\Partder}[1]{\frac{\partial }{\partial #1}}
\newcommand{\Partdern}[2]{\frac{\partial^#1 }{\partial#2 ^#1}}
\newcommand{\partder}[2]{\frac{\partial #1}{\partial #2}}
\newcommand{\partdern}[3]{{\frac{\partial^#1 #2}{\partial #3^#1}}}
\newcommand{\Cov}{\mathrm{Cov}}
\newcommand{\eff}{_{\mathrm{eff}}}
\newcommand{\SEp}{_{\mathrm{SEp}}}
\newcommand{\Irred}{\mathrm{Irred}}
\newcommand{\Sep}{\mathrm{Sep}}
\newcommand{\Nonsep}{\mathrm{Nonsep}}
\newcommand{\Irr}{_{\mathrm{Irr}}}
\newcommand{\RL}{_{\mathrm{L}}}
\newcommand{\QL}{_{\mathrm{QL}}}
\newcommand{\sgn}{\mathrm{sgn}}
\newcommand{\etc}{\mathrm{etc}}
\newcommand{\out}{\mathrm{out}}
\newcommand{\iin}{\mathrm{in}}
\newcommand{\inter}{\mathrm{int}}
\renewcommand{\rm}{\mathrm}
\newcommand{\mI}{\mathrm{I}}
\newcommand{\mS}{\mathrm{S}}
\newcommand{\mD}{\mathrm{D}}
\newcommand{\mG}{\mathrm{G}}
\newcommand{\mH}{\mathrm{H}}
\newcommand{\ext}{\mathrm{ext}}
\newcommand{\conn}{\mathrm{conn}}
\newcommand{\Counter}{\mathrm{Counter}}
\newcommand{\Linked}{\mathrm{linked}}
\newcommand{\linked}{\mathrm{linked}}
\newcommand{\mB}{\mathrm{B}}
\newcommand{\mC}{\mathrm{C}}
\newcommand{\SE}{\mathrm{SE}}
\newcommand{\G}{\mathrm{G}}
\newcommand{\sr}{\mathrm{sr}}
\newcommand{\op}{\mathrm{op}}
\newcommand{\opL}{\mathrm{opL}}
\newcommand{\cl}{\mathrm{cl}}
\newcommand{\clC}{\mathrm{clC}}
\newcommand{\bs}{\boldsymbol}
\newcommand{\bdot}{\boldsymbol\cdot}
\newcommand{\bnabla}{\bs{\nabla}}
\newcommand{\Q}{\mathcal{\boldsymbol{Q}}}
\newcommand{\bOm}{\mathcal{\boldsymbol{\Om}}}
\renewcommand{\H}{H}
\newcommand{\A}{A}
\newcommand{\bH}{\bs{\H}}
\newcommand{\bPsi}{\bs{\Psi}}
\newcommand{\bC}{\bs{C}}
\newcommand{\bR}{\bs{R}}
\newcommand{\bV}{\bs{V}}
\newcommand{\U}{U}
\newcommand{\Uop}{U_\op}
\newcommand{\Ucl}{U_\cl}
\newcommand{\p}{\hat{p}}
\newcommand{\hbp}{\hat{\bs{p}}}
\newcommand{\calH}{{\mathcal{H}}}
\newcommand{\calO}{\mathcal{O}}
\newcommand{\hO}{\hat{\calO}}
\newcommand{\calG}{\mathcal{G}}
\newcommand{\calP}{\mathcal{P}}
\newcommand{\Psit}{\widetilde{\Psi}}
\newcommand{\calGop}{\mathcal{G}_\op}
\newcommand{\calGcl}{\mathcal{G}_\cl}
\newcommand{\calV}{\mathcal{V}}
\newcommand{\calW}{\mathcal{U}}
\newcommand{\calU}{\mathcal{U}}
\newcommand{\calVsf}{\mathcal{V_\rm{sp}}}
\newcommand{\calL}{\mathcal{L}}
\newcommand{\IPair}{I^\rm{Pair}}
\newcommand{\dagg}{^{\dag}}
\newcommand{\intd}[1]{\int\frac{\dif #1}{2\pi}}
\newcommand{\half}{{\displaystyle\frac{1}{2}}}
\newcommand{\halfi}{{\displaystyle\frac{\im}{2}}}
\newcommand{\dint}{\int\!\!\!\int}
\newcommand{\tint}{\int\!\!\!\int\!\!\!\int}
\newcommand{\sint}{\tint\!\!\!\tint}
\newcommand{\ddint}{\dint\!\!\!\dint}
\newcommand{\dintd}[2]{\dint\frac{\dif #1}{2\pi}\,\frac{\dif #2}{2\pi}}
\newcommand{\tintd}[3]{\dint\frac{\dif #1}{2\pi}\,\frac{\dif #2}{2\pi}\,\frac{\dif #3}{2\pi}}
\newcommand{\ddintd}[4]{\ddint\frac{\dif #1}{2\pi}\,\frac{\dif #2}{2\pi}\,
\frac{\dif #3}{2\pi}\,\frac{\dif #4}{2\pi}}
\newcommand{\dddintd}[6]{\sint\frac{\dif #1}{2\pi}\,\frac{\dif #2}{2\pi}\,
\frac{\dif #3}{2\pi}\,\frac{\dif #4}{2\pi}\,\frac{\dif
#5}{2\pi}\,\frac{\dif #6}{2\pi}}
\newcommand{\ddt}{\frac{\partial}{\partial t}}
\newcommand{\intbr}{\int\dif\br\,}
\newcommand{\gamlim}{\ensuremath{\gamma\rightarrow 0}}
\newcommand{\vsp}{\vspace{0.5cm}}
\newcommand{\vvsp}{\vspace{0.25cm}}
\newcommand{\vvvsp}{\vspace{0.125cm}}
\newcommand{\mvsp}{\vspace{-0.5cm}}
\newcommand{\mvvsp}{\vspace{-0.25cm}}
\newcommand{\mvvvsp}{\vspace{-0.25cm}}
\newcommand{\mmvsp}{\vspace{-0.25cm}}
\newcommand{\mVsp}{\vspace{-1cm}}
\newcommand{\mVSP}{\vspace{-2cm}}
\newcommand{\Vsp}{\vspace{1cm}}
\newcommand{\Wsp}{\vspace{2cm}}
\newcommand{\hsp}{\hspace{0.5cm}}
\newcommand{\hhsp}{\hspace{0.25cm}}
\newcommand{\Hsp}{\hspace{1cm}}
\newcommand{\HSP}{\hspace{2cm}}
\newcommand{\mmhsp}{\hspace{-0.25cm}}
\newcommand{\mhsp}{\hspace{-0.5cm}}
\newcommand{\mHsp}{\hspace{-1cm}}
\newcommand{\mHSP}{\hspace{-2cm}}
\newcommand{\ö}{\"{o}}
\newcommand{\š}{\"{o}}
\newcommand{\Ö}{\"{O}}
\newcommand{\…}{\"{O}}
\newcommand{\ä}{\"a}
\newcommand{\å}{\aa}
\newcommand{\Å}{\AA}
\newcommand{\ue}{\"{u}}
\newcommand{\hpsi}{\hat{\psi}}
\newcommand{\wt}[1]{\widetilde{#1}}
\newcommand{\WU}{\widetilde{U}}
\renewcommand{\it}{\textit}
\renewcommand{\bf}{\textbf}
\newcommand{\bfit}[1]{\textbf{\it{#1}}}
\newcommand{\bful}[1]{\textbf{\ul{#1}}}
\newcommand{\ul}{\underline}
\newcommand{\itul}[1]{\it{\ul{#1}}}
\newcommand{\abs}[1]{|{#1}|}
\newcommand{\Abs}[1]{\big|{#1}\big|}
\newcommand{\eq}{\eqref}
\newcommand{\rarr}{\rightarrow}
\newcommand{\larr}{\leftarrow}
\newcommand{\lrarr}{\leftrightarrow}
\newcommand{\LRarr}{\Longleftrightarrow}
\newcommand{\Rarr}{\Rightarrow}
\newcommand{\Lrarr}{\Longrightarrow}
\newcommand{\rr}{r_{12}}
\newcommand{\frr}{\frac{1}{r_{12}}}
\newcommand{\brr}{\br_{12}}
\newcommand{\DF}{D_{\rm{F}\nu\mu}}
\newcommand{\DFS}{D_{\rm{F}}}
\newcommand{\DFmn}{D_{\rm{F}\mu\nu}}
\newcommand{\DFCij}{D^C_{\rm{Fij}}}
\newcommand{\SF}{S_{\rm{F}}}
\newcommand{\hSF}{\hat{S}_{\rm{F}}}
\newcommand{\hh}{\hat{h}}
\newcommand{\QED}{\rm{QED}}
\newcommand{\CQED}{\rm{CQED}}
\newcommand{\DQED}{\rm{DQED}}
\newcommand{\ren}{\rm{ren}}
\newcommand{\bou}{\rm{bou}}
\newcommand{\free}{\rm{free}}
\renewcommand{\sp}[2]{\bra{#1}{#2}\rangle}
\newcommand{\SP}[2]{\Bigbra{#1}{#2}\Big\rangle}
\newcommand{\qand}{\quad\rm{and}\quad}
\newcommand{\V}{V}
\newcommand{\VC}{V_\rm{C}}
\newcommand{\VF}{V_\rm{F}}
\newcommand{\VG}{V_\rm{G}}
\newcommand{\VspC}{V_\rm{TC}}
\newcommand{\Vsr}{V_\rm{sr}}
\newcommand{\MC}{\calM_\rm{C}}
\newcommand{\IC}{I^\rm{C}}
\newcommand{\ICC}{I^\rm{C}_\rm{C}}
\newcommand{\ICT}{I^\rm{C}_\rm{T}}
\newcommand{\UT}{U_\rm{T}}
\newcommand{\VT}{V_\rm{T}}
\newcommand{\vT}{v_\rm{T}}
\newcommand{\HD}{H_\rm{D}}
\newcommand{\fC}{f^\rm{C}}
\newcommand{\fCC}{f^\rm{C}_\rm{C}}
\newcommand{\fCT}{f^\rm{C}_\rm{T}}
\newcommand{\fF}{f^\rm{F}}
\newcommand{\VB}{V_\rm{B}}
\newcommand{\Vsf}{V_\rm{sp}}
\newcommand{\Vgsf}{\calV_\rm{T}}
\newcommand{\Vsfp}{V_\rm{sp}}
\newcommand{\Usfp}{U_\rm{sp}}
\newcommand{\Msfp}{\calM_\rm{sp}}
\newcommand{\MVx}{\calM_\rm{Vx}}
\newcommand{\VSE}{V_\rm{SE}}
\newcommand{\VVert}{V_\rm{Vx}}
\newcommand{\VVx}{V_\rm{Vx}}
\newcommand{\VspGen}{V_\rm{tr}^\rm{Gen}}
\newcommand{\VTC}{V_\rm{TC}}
\newcommand{\calUsp}{\calU_\rm{sp}}
\newcommand{\Usf}{U_\rm{sp}}
\newcommand{\Ssf}{S_\rm{T}}
\newcommand{\Msf}{\calM_\rm{sp}}
\newcommand{\MT}{\calM_\rm{T}}
\newcommand{\Ksf}{{\cal K_\rm{sp}}}
\newcommand{\MSE}{\calM_\rm{SE}}
\newcommand{\SSE}{S_\rm{SE}}
\newcommand{\USE}{U_\rm{SE}}
\newcommand{\Coul}{\rm{Coul}}
\newcommand{\calK}{\mathcal{\kappa}}
\newcommand{\calKc}{I_c}
\newcommand{\calF}{\mathcal{F}}
\newcommand{\calM}{{\mathcal{M}}}
\newcommand{\calR}{\mathcal{\hat{R}}}
\newcommand{\calB}{\mathcal{B}}
\newcommand{\GQ}{\Gamma_Q}
\newcommand{\GV}{\Gamma V}
\newcommand{\Gam}{\Gamma}
\newcommand{\GQV}{\GQ V}
\newcommand{\GI}{\calG^\rm{I}}
\newcommand{\bGQ}{\bs{\Gamma_Q}}
\newcommand{\Gk}{\bs{\Gamma_Q}}
\newcommand{\Gv}{\Gamma_Q}
\newcommand{\TD}{T_\rm{D}}
\newcommand{\F}{\rm{F}}
\renewcommand{\P}{\bs{P}}
\newcommand{\Util}{\widetilde{U}}
\newcommand{\Ucov}{U_\Cov}
\renewcommand{\S}{S}
\newcommand{\Vtil}{\widetilde{V}}
\newcommand{\Om}{\Omega}
\newcommand{\Omsf}{\Omega_\rm{sp}}
\newcommand{\Omi}{\Om_\rm{I}}
\newcommand{\OmI}{{\Om_\rm{I}}}
\newcommand{\om}{\omega}
\newcommand{\OM}{\mathcal{\boldsymbol{\Om}}}
\newcommand{\Ombar}{\bar{\Omega}}
\newcommand{\Vbar}{\bar{V}}
\newcommand{\VI}{V_{12}}
\newcommand{\VR}{V_{\rm{R}}}
\newcommand{\VRbar}{\Bar{V}_{\rm{R}}}
\newcommand{\calE}{{\mathcal{E}}}
\newcommand{\CALE}{{\mathcal{E}}}
\newcommand{\la}{\lambda}
\newcommand{\FLL}{F_\rm{LL}}
\newcommand{\ka}{|\bs{k}|}
\newcommand{\norm}[1]{||{#1}||}
\newcommand{\Norm}[1]{\big|\big|{#1}\big|\big|}
\newcommand{\NORM}[1]{\Big|\Big|{#1}\Big|\Big|}
\newcommand{\Fr}{Fr\'echet }
\newcommand{\Ga}{G\^ateaux }
\newcommand{\epsn}{\ensuremath{\epsilon}}
\newcommand{\Der}[1]{\frac{\dif}{\dif#1}}
\newcommand{\der}[2]{\frac{\dif#1}{\dif#2}}
\newcommand{\ave}[1]{\langle #1\rangle}
\newcommand{\Ave}[1]{\Big\langle #1\Big\rangle}
\newcommand{\BB}[2]{\Big\{{#1}\,\Big|\:{#2}\Big\}}
\newcommand{\LL}{L^1\cap L^3}
\newcommand{\partdelta}[2]{\frac{\delta #1}{\delta #2}}
\newcommand{\pd}[1]{\frac{\delta #1}{\delta \calE}}
\newcommand{\pda}[1]{\frac{\delta^* #1}{\delta \calE}}
\newcommand{\Pda}[1]{\frac{\delta^*}{\delta #1}}
\newcommand{\Partdelta}[1]{\frac{\delta }{\delta #1}}
\newcommand{\Pd}[1]{\frac{\delta }{\delta #1}}
\newcommand{\partdeltan}[3]{\frac{\delta^#1 #2}{\delta #3^#1}}
\newcommand{\partdeltanp}[3]{\frac{\delta^{(#1)} #2}{\delta
#3^{(#1)}}}
\newcommand{\pdna}[2]{\frac{\delta^{^*#1} #2}{\delta\calE^#1}}
\newcommand{\pdn}[2]{\frac{\delta^{#1} #2}{\delta\calE^{#1}}}
\newcommand{\pdnp}[2]{\frac{\delta^{(#1)} #2}{\delta\calE^{(#1)}}}
 \newcommand{\ett}{^{(1)}}
\newcommand{\nol}{^{(0)}}
\newcommand{\tva}{^{(2)}}
\newcommand{\tre}{^{(3)}}
\newcommand{\fyr}{^{(4)}}
\newcommand{\enh}{^{(1/2)}}
\newcommand{\treh}{^{(3/2)}}
\newcommand{\femh}{^{(5/2)}}
\newcommand{\n}{^{(n)}}
\newcommand{\m}{^{(m)}}
\newcommand{\nm}{^{(n-m)}}
\newcommand{\nett}{^{(n-1)}}
\newcommand{\ntva}{^{(n-2)}}
\newcommand{\PE}{P_\mathcal{E}}
\newcommand{\GE}{\Gamma(\calE)}
\newcommand{\GEE}{\Gamma(E)}
\newcommand{\GQEE}{\Gamma_Q(E)}
\newcommand{\GQE}{\Gamma_Q(\calE)}
\newcommand{\GQEP}{\Gamma_Q(\EP)}
\newcommand{\GEN}{\Gamma(E_0)}
\newcommand{\GQEN}{\Gamma_Q(E_0)}
\newcommand{\GEP}{\Gamma(\calE')}
\newcommand{\GaV}{\Gamma V}
\newcommand{\GaVg}{\GQ V}
\newcommand{\PEP}{P_{\calE'}}
\newcommand{\QE}{Q_{\calE}}
\newcommand{\QEP}{Q_{\calE'}}
\newcommand{\PEPP}{P_{\calE''}}
\newcommand{\EP}{\calE'}
 \newcommand{\limgam}{\lim_{\gamlim}}
 \newcommand{\Ugam}[1]{U_\gamma(#1,-\infty)}
\newcommand{\Ugamtil}[1]{\widetilde{U}_\gamma(#1,-\infty)}
\newcommand{\Ugamt}{\widetilde{U}_\gamma}
\newcommand{\img}{\im\gamma}
\newcommand{\ime}{\im\eta}
\newcommand{\pbar}{\not\!p}
\newcommand{\psl}{\!\not\!p\,}
\newcommand{\Asl}{\not\!\! A\,}
\newcommand{\qsl}{\not\!q}
\newcommand{\ksl}{\not\!k}
\newcommand{\lsl}{\not\!l}
\newcommand{\nott}{\not\!\!}
\newcommand{\wtp}{\widetilde{p}}
\newcommand{\wtq}{\widetilde{q}}
\newcommand{\wtk}{\widetilde{k}}
\newcommand{\Deltag}{\Delta_\gamma}
\newcommand{\Deltatg}{\Delta_{2\gamma}}
\newcommand{\Deltafg}{\Delta_{4\gamma}}
\renewcommand{\bar}{\setlength{\unitlength}{0.6cm}\put(0,0.6){\line(1,0){0.4}}}
\newcommand{\fbar}{\setlength{\unitlength}{0.6cm}\put(0,0.4){\line(1,0){0.4}}}
\newcommand{\MSC}{\rm{MSC}}
\newcommand{\IF}{\rm{IF}}
                                                  \newcommand{\Htil}{\widetilde{H}}
                                                  \newcommand{\Ubar}{\bar{U}}
                                                  \newcommand{\Hbar}{\bar{V}}
                                                  \newcommand{\Udot}{\dot{U}}
                                                  \newcommand{\Ubardot}{\dot{\Ubar}}
                                                  \newcommand{\Utildot}{\dot{\Util}}
                                                  \newcommand{\Cdot}{\dot{C}}
                                                  \newcommand{\Ombardot}{\dot{\Ombar}}
                                                  \newcommand{\bsdot}{\bs{\cdot}}

\newcommand{\npartdelta}[3]{\frac{\delta^#1 #2}{\delta #3^#1}}
\newcommand{\ip}[1]{| #1 \rangle \langle #1 |}
\newcommand{\con}{\mathrm{con}}
\newcommand{\ph}{\mathrm{ph}}
\newcommand{\bA}{\bs{A}}
\newcommand{\Pmu}{\partial^\mu}
\newcommand{\Pnu}{\partial^\nu}
\newcommand{\pmu}{\partial_\mu}
\newcommand{\pnu}{\partial_\nu}
\newcommand{\bAT}{\bs{A}_\perp}
\newcommand{\bAL}{\bs{A}_\parallel}
\newcommand{\bET}{\bs{E}_\perp}
\newcommand{\bEL}{\bs{E}_\parallel}
\newcommand{\bB}{\bs{\rm{B}}}
\newcommand{\bE}{\bs{\rm{E}}}
\newcommand{\bn}{\bs{n}}
\newcommand{\beps}{\bs{\eps}}
\newcommand{\HI}{H_\rm{int,I}}
\newcommand{\calHI}{{\cal H}_\rm{int,I}}
\newcommand{\ih}{\frac{\im}{\hbar}}
\newcommand{\bkx}{\bk\bdot\bx}
\newcommand{\halfS}{{\textstyle\frac{1}{2}\,}}
\newcommand{\h}{\hat{h}}
\newcommand{\DFnu}{D_{\rm{F}\nu\mu}}
\newcommand{\Eab}{E_{ab}}
\newcommand{\rE}{{\red E}}
\newcommand{\sphline}{\hline\vsp}
\newcommand{\Ram}[4]
{\begin{picture}(0,0)(0,0)\setlength{\unitlength}{1cm}
    \put(#1,#2){\Ebox{#3}{#4}}
  \end{picture}}
                                                  \newcommand{\Ers}{E_{rs}}
                                                  \newcommand{\Etu}{E_{tu}}
                                                  \newcommand{\Eru}{E_{ru}}
                                                  \newcommand{\Ecd}{E_{cd}}
                                                  \newcommand{\Erd}{E_{rd}}
                                                  \newcommand{\epsi}{\epsilon}

%% file: FigurecommandsV.tex

\newcommand{\NVP}
{\put(0,0){\LineV{4}}\put(2,0){\LineV{4}}}
\newcommand{\SVP}
{\put(0,0){\LineV{2.5}}\put(-1,1.5){\LineV{2.5}}\put(0,2.5){\LineDl{}}
\put(2,0){\LineV{4}}}
\newcommand{\DVP}
{\put(0,0){\LineV{2.5}}\put(-1,1.5){\LineV{2.5}}\put(0,2.5){\LineDl{}}
\put(2,0){\LineV{2.5}}\put(3,1.5){\LineV{2.5}}\put(2,2.5){\LineDr{}}}
\newcommand{\SVPC}
{\put(0,0){\LineV{3}}\put(-1,2){\LineV{2}}\put(0,3){\LineDl{}}
\put(2,0){\LineV{4}}}
\newcommand{\DVPC}
{\put(0,0){\LineV{3}}\put(-1,2){\LineV{2}}\put(0,3){\LineDl{}}
\put(2,1){\LineV{3}}\put(3,0){\LineV{2}}\put(3,2){\LineDl{}}}

\newcommand{\circl}[0]
{\circle*{0.225} }

\newcommand{\LineH}[1]
{\linethickness{0.5mm} \put(0,0){\line(1,0){#1}} }

\newcommand{\LineHH}[1]
{\linethickness{1mm} \put(0,0){\line(1,0){#1}} }

\newcommand{\LineD}[1]
{\put(0,-0.2){\line(1,0){#1}} \put(0,0.2){\line(1,0){#1}}}

\newcommand{\LineDD}[1]
{\put(0,-0.3){\line(1,0){#1}} \put(0,0.3){\line(1,0){#1}}}

\newcommand{\LineS}[1]
{\linethickness{1mm}
\put(0,0){\line(1,0){#1}}
}

\newcommand{\LineWO}[1]
{\linethickness{0.75mm} \put(0,0){\line(1,0){#1}} }

\newcommand{\LineV}[1]
{\linethickness{0.5mm}  \put(0,0){\line(0,1){#1}} }

\newcommand{\LineEtt}[2]
{\linethickness{0.5mm}  \put(0,0){\line(0,1){#1}}
\put(0,#2){\VectorUp}}

\newcommand{\LineTva}[3]
{\linethickness{0.5mm}  \put(0,0){\line(0,1){#1}}
\put(0,#2){\VectorUp}\put(0,#3){\VectorUp} }

\newcommand{\LineTre}[4]
{\linethickness{0.5mm}  \put(0,0){\line(0,1){#1}}
\put(0,#2){\VectorUp}\put(0,#3){\VectorUp}\put(0,#4) {\VectorUp}}

\newcommand{\LineFyr}[5]
{\linethickness{0.5mm}  \put(0,0){\line(0,1){#1}}
\put(0,#2){\VectorUp}\put(0,#3){\VectorUp}\put(0,#4){\VectorUp}\put(0,#5)
{\VectorUp} }

\newcommand{\LineFem}[6]
{\linethickness{0.5mm}  \put(0,0){\line(0,1){#1}}
\put(0,#2){\VectorUp}\put(0,#3){\VectorUp}\put(0,#4){\VectorUp}\put(0,#5)
{\VectorUp}\put(0,#6){\VectorUp} }

\newcommand{\LineSex}[7]
{\linethickness{0.5mm}  \put(0,0){\line(0,1){#1}} 
\put(0,#2){\VectorUp}\put(0,#3){\VectorUp}\put(0,#4){\VectorUp}\put(0,#5)
{\VectorUp}\put(0,#6){\VectorUp}\put(0,#7){\VectorUp}
}

\newcommand{\LineVT}[1]
{\linethickness{0.3mm}  \put(0,0){\line(0,1){#1}} }

\newcommand{\Linev}[1]
{\put(0,0){\line(0,1){#1}} }

\newcommand{\LineW}[1]
{\linethickness{0.75mm} \put(0,0){\line(0,1){#1}}}

\newcommand{\LineHpt}[1]
{\put(0,0.015){\line(1,0){#1}} \put(0,-0.015){\line(1,0){#1}}
\put(0,0){\circl} \put(#1,0){\circl}}

\newcommand{\LineDl}[1]
{\put(0.012,-0.012){\line(-1,-1){#1}}\put(-0.012,0.012){\line(-1,-1){#1}}
\put(0.0,0.0){\line(-1,-1){#1}}}

\newcommand{\LineUl}[1]
{\put(0.012,-0.012){\line(-1,1){#1}}\put(-0.012,0.012){\line(-1,1){#1}}
\put(0.0,0.0){\line(-1,1){#1}}}

\newcommand{\Lineul}[1]
{\put(0.012,-0.012){\line(-1,2){#1}}\put(-0.012,0.012){\line(-1,2){#1}}
\put(0.0,0.0){\line(-1,2){#1}}}

\newcommand{\Linedl}[1]
{\put(0.01,0.01){\line(-1,-2){#1}}
\put(-0.01,-0.01){\line(-1,-2){#1}}}

\newcommand{\LineUr}[1]
{\put(0,0.015){\line(1,1){#1}}
\put(0,0){\line(1,1){#1}}
\put(0,-0.015){\line(1,1){#1}}}

\newcommand{\LineDr}[1]
{\put(0,0.01){\line(1,-1){#1}} \put(0,0){\line(1,-1){#1}}
\put(0,-0.01){\line(1,-1){#1}}}

\newcommand{\Linedr}[1]
{\put(0,0.01){\line(1,-2){#1}} \put(0,0){\line(1,-2){#1}}
\put(0,-0.01){\line(1,-2){#1}}}

\newcommand{\Lineur}[1]
{\put(0,0.025){\line(1,2){#1}} \put(0,0){\line(1,2){#1}}
\put(0,-0.025){\line(1,2){#1}}}

\newcommand{\DLine}[1]
{\put(0,-0.05){\line(1,0){#1}} \put(0,0.05){\line(1,0){#1}}
\put(0,0){\circl}}

\newcommand{\Vector}[0]
{\thicklines\setlength{\unitlength}{1cm}\put(-0.10,-0.035){\vector(-1,0){0}}}

\newcommand{\VectorR}[0]
{\thicklines\setlength{\unitlength}{1cm}\put(0.14,0.035){\vector(1,0){0}}}

\newcommand{\vectorR}[0]
{\setlength{\unitlength}{1cm}\put(0.13,0.1){\vector(1,0){0}}}

\newcommand{\VectorUp}[0]
{\thicklines\setlength{\unitlength}{1cm}
\put(0,0.15){\vector(0,0){0}}}

\newcommand{\vectorUp}[0]
{\setlength{\unitlength}{1cm} \put(0,0.12){\vector(0,0){0}} }

\newcommand{\VectorT}[0]
{\thicklines\setlength{\unitlength}{1cm}\put(0,0.18){\vector(0,0){0}}
\put(0.02,0.02){\vector(0,0){0}}}

\newcommand{\VectorDn}[0]
{\thicklines\setlength{\unitlength}{1cm}
\put(0,-0.15){\vector(0,-1){0}}}

\newcommand{\VectorDl}[0]
{\thicklines \setlength{\unitlength}{1cm}
\put(-0.1,-0.1){\vector(-1,-1){0}}}

\newcommand{\VectorDr}[0]
{\thicklines\setlength{\unitlength}{1cm}
\put(0.084,-0.092){\vector(1,-1){0}}}

\newcommand{\Vectordr}[0]
{\thicklines\setlength{\unitlength}{1cm}
\put(0.022,0.112){\vector(1,-3){0}} }

\newcommand{\Vectorur}[0]
{\thicklines\setlength{\unitlength}{1cm}
\put(0.04,-0.062){\vector(1,2){0}} }

\newcommand{\VectorUr}[0]
{\thicklines\setlength{\unitlength}{1cm}
 \put(0.22,-0.1){\vector(1,1){0}}}

\newcommand{\VectorUl}[0]
{\put(-0.23,-0.02){\vector(-1,1){0}}
\put(-0.19,-0.03){\vector(-1,1){0}}
\put(-0.22,-0.06){\vector(-1,1){0}} }

\newcommand{\Vectorul}[0]
{\put(-0.13,-0.02){\vector(-1,2){0}}
\put(-0.11,-0.04){\vector(-1,2){0}}
\put(-0.16,-0.06){\vector(-1,2){0}}}

\newcommand{\Wector}[0]
{\put(-0.15,0){\Vector}\put(0.15,0){\Vector}}

\newcommand{\WectorUp}[0]
{\put(0,0.125)\VectorUp\put(0,-0.125)\VectorUp}

\newcommand{\WectorDn}[0]
{\put(0,0.125)\VectorDn\put(0,-0.125)\VectorDn}

\newcommand{\WectorDl}[0]
{\put(0.1,0.1)\VectorDl\put(-0.1,-0.1)\VectorDl}

\newcommand{\Wectordl}[0]
{\setlength{\unitlength}{1cm}
 \put(0.04,0.10){\vector(-2,-1){0}}\put(0.02,0.13){\vector(-2,-1){0}}
 \put(-0.14,0.02){\vector(-2,-1){0}}\put(-0.16,0.05){\vector(-2,-1){0}}}

\newcommand{\EllineH}[4]
{\put(0,0){\LineH{#1}} \put(#2,0){\Vector}
\put(#2,0.45){\makebox(0,0){$#3$}}
\put(#2,-0.35){\makebox(0,0){$#4$}}}

\newcommand{\lline}[4]
{\put(0,0){\LineV{#1}} \put(-0.3,#2){\makebox(0,0){$#3$}}
\put(0.4,#2){\makebox(0,0){$#4$}}}

\newcommand{\Elline}[4]
{\put(0,0){\LineV{#1}} \put(0,#2){\VectorUp}
\put(-0.35,#2){\makebox(0,0){$#3$}}
\put(0.4,#2){\makebox(0,0){$#4$}} }

\newcommand{\EllineDV}[3]
{\put(0,0){\LineV{#1}}\put(0,#2){\VectorUp} \put(0,#3){\VectorUp}}

\newcommand{\EllineTV}[4]
{\put(0,0){\LineV{#1}}\put(0,#2){\VectorUp}
\put(0,#3){\VectorUp}\put(0,#4){\VectorUp}}

\newcommand{\EllineFV}[5]
{\put(0,0){\LineV{#1}}\put(0,#2){\VectorUp}
\put(0,#3){\VectorUp}\put(0,#4){\VectorUp}\put(0,#5){\VectorUp}}

\newcommand{\EllineSV}[6]
{\put(0,0){\LineV{#1}}\put(0,#2){\VectorUp}
\put(0,#3){\VectorUp}\put(0,#4){\VectorUp}\put(0,#5){\VectorUp}
\put(0,#6){\VectorUp}}

\newcommand{\elline}[3]
{\put(0,0){\LineV{#1}} \put(0,0){\makebox(0,#1){\VectorUp}}
\put(-0.4,0){\makebox(0,#1){$#2$}}
\put(0.5,0){\makebox(0,#1){$#3$}}
}

\newcommand{\ellineP}[6]
{\put(0,0){\elline{#1}{#3}{#4}}\put(#2,0){\elline{#1}{#5}{#6}} 
}

\newcommand{\EllineW}[4]
{\put(0,0){\LineW{#1}} \put(0,#2){\WectorUp}
\put(-0.4,#2){\makebox(0,0){$#3$}}
\put(0.5,#2){\makebox(0,0){$#4$}} }

\newcommand{\Ellinev}[4]
{\put(0,0){\Linev{#1}} \put(0,#2){\VectorUp}
\put(-0.3,#2){\makebox(0,0){$#3$}}
\put(0.4,#2){\makebox(0,0){$#4$}} }

\newcommand{\DElline}[4]
{\put(0,0){\LineV{#1}} \put(0,#2){\WectorUp}
\put(-0.3,#2){\makebox(0,0){$#3$}}
\put(0.3,#2){\makebox(0,0){$#4$}}}

\newcommand{\DEllineDn}[4]
{\put(0,0){\LineV{#1}} \put(0,#2){\WectorDn}
\put(-0.25,#2){\makebox(0,0){$#3$}}
\put(0.25,#2){\makebox(0,0){$#4$}}}

\newcommand{\Ellinet}[4]
{\put(0,0){\Linev{#1}} \put(0,#2){\vector(0,1){0}}
\put(-0.35,#2){\makebox(0,0){$#3$}}
\put(0.35,#2){\makebox(0,0){$#4$}}}

\newcommand{\EllineT}[4]
{\put(0,0){\LineW{#1}} \put(0,#2){\VectorUp}
\put(-0.35,#2){\makebox(0,0){$#3$}}
\put(0.35,#2){\makebox(0,0){$#4$}}}

\newcommand{\EllineDnt}[4]
{\put(0,0){\Linev{#1}} \put(0,#2){\VectorDn}
\put(-0.35,#2){\makebox(0,0){$#3$}}
\put(0.35,#2){\makebox(0,0){$#4$}}}

\newcommand{\EllineDn}[4]
{\put(0,0){\LineV{#1}} \put(0,#2){\VectorDn}
\put(-0.35,#2){\makebox(0,0){$#3$}}
\put(0.5,#2){\makebox(0,0){$#4$}}}

\newcommand{\EllineDl}[4]
{\put(0,0){\LineDl{#1}} \put(-#2,-#2){\VectorDl}
\put(-#2,-#2){\makebox(-0.25,0.25){$#3$}}
\put(-#2,-#2){\makebox(0.5,-0.25){$#4$}}}

\newcommand{\Ellinedl}[4]
{\put(0.01,0.01){\line(-1,-2){#1}}
\put(-0.01,-0.01){\line(-1,-2){#1}}
\thicklines\put(-0.05,-0.1){\vector(-1,-2){#2}}
\put(-0.4,-0.4){\makebox(-#1,-#1){$#3$}}
\put(-0.5,-0.6){\makebox(-#1,-#1){$#4$}}}

\newcommand{\EllinedL}[4]
{\put(0.02,0.02){\line(-1,-3){#1}}
\put(-0.02,-0.02){\line(-1,-3){#1}}
\thicklines\put(-0.05,-0.1){\vector(-1,-3){#2}}
\put(-0.4,0){\makebox(-1,-3){$#3$}}
\put(0.4,-0){\makebox(-1,-3){$#4$}}}

\newcommand{\Ellinedr}[4]
{\put(0.01,0.01){\line(1,-2){#1}}
\put(-0.01,-0.01){\line(1,-2){#1}}
\thicklines\put(0.05,-0.1){\vector(1,-2){#2}}
\put(-0.4,0){\makebox(1,-2){$#3$}}
\put(0.2,0.2){\makebox(1,-2){$#4$}}}

\newcommand{\EllineA}[7]
{\put(0.0,0.0){\line(#1,#2){#3}} \put(0.005,0.0){\line(#1,#2){#3}}
\put(-0.005,0.0){\line(#1,#2){#3}} \put(0,0){\vector(#1,#2){#4}}
\put(0.010,0){\vector(#1,#2){#4}}
\put(-0.010,0){\vector(#1,#2){#4}}
\put(#6,#7){\makebox(0,0){$#5$}}}

\newcommand{\EllineDr}[4]
{\put(0,0){\LineDr{#1}} \put(#2,-#2){\makebox(0.2,0.2)\VectorDr}
\put(#2,-#2){\makebox(-0.4,-0.3){$#3$}}
\put(#2,-#2){\makebox(0.75,0.25){$#4$}}}

\newcommand{\EllinedR}[5]
{\put(0,0){\line(1,-3){#1}} \put(0.014,0){\line(1,-3){#1}}
\put(-0.014,0){\line(1,-3){#1}}
\put(#2,-#3){\makebox(0,0){{\Vectordr}}}
\put(#2,-#3){\makebox(-0.5,-0.5){$#4$}}
\put(#2,-#3){\makebox(0.5,0.5){$#5$}}}

\newcommand{\EllineuR}[5]
{\put(0,0){\line(1,3){#1}} \put(0.014,0){\line(1,3){#1}}
\put(-0.014,0){\line(1,3){#1}}
\put(#2,#3){\makebox(0,0){\Vectorur}}
\put(#2,#3){\makebox(-0.5,-0.5){$#4$}}
\put(#2,#3){\makebox(0.5,0.5){$#5$}}}

\newcommand{\EllineUl}[4]
{\put(0,0){\LineUl{#1}} \put(-#2,#2){{\makebox(0,-0.15)\VectorUl}}
\put(-#2,#2){\makebox(-0.5,0.5){$#3$}}
\put(-#2,#2){\makebox(0.5,0.5){$#4$}}}

\newcommand{\Ellineul}[5]
{\put(0,0){\Lineul{#1}} \put(#2,#3){\makebox(0.5,0){\Vectorul}}
\put(#2,#3){\makebox(-0.5,0){$#4$}}\put(#2,#3){\makebox(0.5,0){$#5$}}
}

\newcommand{\Ellineur}[5]
{\put(0,0){\Lineur{#1}}
\put(#2,#3){\makebox(0,0){\Vectorur}}
\put(#2,#3){\makebox(-0.5,0){$#4$}}
\put(#2,#3){\makebox(0.5,0){$#5$}}}

\newcommand{\EllineUr}[4]
{\put(0,0){\LineUr{#1}} \put(#2,#2){\makebox(-0.35,0){\VectorUr}}
\put(-0.2,0.4){\makebox(#1,#1){$#3$}}
\put(0.2,-0.1){\makebox(#1,#1){$#4$}}}

\newcommand{\DEllineDl}[4]
{\put(0,0){\LineDl{#1}}
\put(-#2,-#2){\WectorDl}
\put(-0.25,0.25){\makebox(-#1,-#1){$#3$}}
\put(0.25,-0.25){\makebox(-#1,-#1){$#4$}}}

\newcommand{\Ebox}[2]
{\put(0,0){\LineH{#1}} \put(0,#2){\LineH{#1}}
\put(0,0){\LineV{#2}} \put(#1,0){\LineV{#2}}}

\newcommand{\EEbox}[2]
{\put(0,0){\LineHH{#1}} \put(0,#2){\LineHH{#1}}
\put(0,0){\LineW{#2}} \put(#1,0){\LineW{#2}}}

\newcommand{\dashH}
{\multiput(0.05,0)(0.25,0){5}{\line(1,0){0.15}}}

\newcommand{\dash}[1]
{\multiput(0.05,0)(0.25,0){#1}{\line(1,0){0.15}}}

\newcommand{\dashV}[1]
{\multiput(0.05,0)(0,0.25){#1}{\line(0,1){0.15}}}

\newcommand{\dashHp}
{\multiput(0.05,0)(0.25,0){6}{\line(1,0){0.15}}}

\newcommand{\DashH}
{\multiput(0.05,0)(0.25,0){10}{\line(1,0){0.15}}}

\newcommand{\dashHnum}[2]
{\multiput(0.05,0)(0.25,0){5}{\line(1,0){0.15}}
\put(-0.25,0){\makebox(0,0){$#1$}}
\put(1.5,0){\makebox(0,0){$#2$}}}

\newcommand{\dashHnuma}[2]
{\multiput(0.05,0)(0.25,0){5}{\line(1,0){0.15}}
\put(0.25,0.25){\makebox(0,0){$#1$}}
\put(1,0.25){\makebox(0,0){$#2$}}}

\newcommand{\dashHnumu}[2]
{\multiput(0.05,0)(0.25,0){5}{\line(1,0){0.15}}
\put(0.25,-0.25){\makebox(0,0){$#1$}}
\put(1,-0.25){\makebox(0,0){$#2$}}}

\newcommand{\Potint}
{\put(0,0)\dashH \put(1.35,0){\makebox(0,0){$\times$}}
\put(0,0){\circle*{0.15}} }

\newcommand{\potint}
{\multiput(0.05,0)(0.25,0){3}{\line(1,0){0.15}}
\put(0.85,0){\makebox(0,0){$\times$}} \put(0,0){\circle*{0.15}}}

\newcommand{\PotintS}
{\put(0,0){\dash{4}} \put(1,0){\makebox(0,0){$\times$}}
\put(0,0){\circle*{0.1}}}

\newcommand{\PotintL}
{\put(-1.25,0)\dashH
\put(-1.35,0){\makebox(0,0){x}}
\put(0,0){\circle*{0.15}}}

\newcommand{\potintL}
{\multiput(0.05,0)(0.25,0){3}{\line(-1,0){0.15}}
\put(-0.85,0){\makebox(0,0){x}} \put(0,0){\circle*{0.15}}}

\newcommand{\Effpot}
{\put(0,0)\dashH \put(1.35,0){\makebox(0,0){$\times$}}
\put(1.35,0){\Circle{0.4}}
}

\newcommand{\EffpotQED}
{\linethickness{0.3mm}\put(0,0)\dashH
\put(1.35,0){\makebox(0,0){x}}
\put(1.35,0){\circle{0.4}}\put(1.35,0){\circle{0.45}}\put(1.35,0){\circle{0.5}}
\put(1.35,0){\circle{0.55}}\put(1.35,0){\circle{0.6}}
\put(0,0){\circle*{0.20}} }

\newcommand{\effpot}
{\multiput(0.05,0)(0.25,0){3}{\line(1,0){0.15}}
\put(0.85,0){\makebox(0,0){x}} \put(0.85,0){\Circle{0.3}}
\put(0,0){\Circle*{0.1}}}

\newcommand{\EffpotL}
{\put(-1.25,0)\dashH \put(-1.25,0){\makebox(0,0){x}}
\put(-1.25,0){\circle{0.4}}
}

\newcommand{\TriangUp}[1]
{\put(-0.7,0){\line(1,0){1.4}} \put(0,0.7){\line(-1,-1){0.7}}
\put(0,0.7){\line(1,-1){0.7}} \put(0,0.25){\makebox(0,0){#1}}}

\newcommand{\TriangDn}[1]
{\put(-0.7,0){\line(1,0){1.4}} \put(0,-0.7){\line(1,1){0.7}}
\put(0,-0.7){\line(-1,+1){0.7}} \put(0,-0.25){\makebox(0,0){#1}}}

\newcommand{\Triang}
{\put(0,0){\line(2,1){0.5}}
\put(0,0){\line(2,-1){0.5}}
\put(0.5,-0.25){\line(0,1){0.5}}}

\newcommand{\Triangle}
{\put(0,0){\line(2,1){0.7}} \put(0,0){\line(2,-1){0.7}}
\put(0.7,-0.35){\line(0,1){0.7}}}

\newcommand{\TriangL}
{\put(0,0){\line(-2,1){0.5}}
\put(0,0){\line(-2,-1){0.5}}
\put(-0.5,-0.25){\line(0,1){0.5}}}

\newcommand{\hfint}
{\put(0,0)\dashH
\put(1.25,0){\makebox(0,0){\Triang}}
\put(0,0){\circle*{0.15}}}

\newcommand{\hfintL}
{\put(-1.25,0)\dashH \put(-1.25,0){\makebox(0,0){\TriangL}}
\put(0,0){\circle*{0.15}}}

\newcommand{\Oval}[2]
{\put(0.0,0){\oval(#1,#2)}
\put(0.04,0){\oval(#1,#2)}\put(-0.04,0){\oval(#1,#2)}
\put(0.0,0.04){\oval(#1,#2)}\put(0.0,-0.04){\oval(#1,#2)}
\put(-0.02,0){\oval(#1,#2)}\put(0.02,0){\oval(#1,#2)}
\put(0,0.02){\oval(#1,#2)}\put(0,0,02){\oval(#1,#2)} }

\newcommand{\LoopT}[1]
{\put(0,0){\circle{1}}\put(0,0.02){\circle{1}}\put(0,-0.025){\circle{1}}
\put(-0.02,0){\circle{1}}\put(0.02,0){\circle{1}} 
\put(0.85,0){\makebox(0,0){#1}}}

\newcommand{\LoopTh}[2]
{\put(0,0){\circle{#1}}\put(0,0.02){\circle{#1}}\put(0,-0.025){\circle{#1}}
\put(-0.02,0){\circle{#1}}\put(0.02,0){\circle{#1}} } 

\newcommand{\VPloop}[2]
{\linethickness{0.4mm}\put(0,0){\LoopTh{#1}{#2}}
 \put(0.5,0.05){\VectorDn}
\put(0.85,0){\makebox(0,0){#2}}}

\newcommand{\VPloopt}[1]
{\put(0,0){\circle{1}} \put(0.44,0){\VectorDn}
\put(0.75,0){\makebox(0,0){$#1$}}}

\newcommand{\VPloopL}[1]
{\linethickness{0.4mm}\put(0,0){\LoopT{#1}}
\put(-0.5,-0.05){\VectorUp} \put(-0.75,0){\makebox(0,0){$#1$}}}

\newcommand{\VPloopLt}[1]
{\put(0,0){\circle{1}} \put(-0.46,-0.05){\VectorUp}
\put(-0.75,0){\makebox(0,0){$#1$}}}

\newcommand{\VPloopLR}[2]
{\linethickness{0,3mm}\put(0,0){\LoopT{#1}} \put(-0.5,0){\VectorDn}
\put(0.5,0){\VectorUp} \put(-0.75,0){\makebox(0,0){$#1$}}
\put(0.75,0){\makebox(0,0){$#2$}}}

\newcommand{\VPloopLRt}[2]
{\put(0,0){\circle{1}}
\put(-0.5,0){\VectorDn}
\put(0.5,0){\VectorUp}
\put(-0.75,0){\makebox(0,0){$#1$}}
\put(0.75,0){\makebox(0,0){$#2$}}}

\newcommand{\VPloopD}[2]
{\linethickness{0,3mm}\put(0,0){\LoopT{#1}} \put(0,0.42){\VectorR}
\put(0,-0.42){\Vector} \put(0,0.8){\makebox(0,0){$#1$}}
\put(0,-0.8){\makebox(0,0){$#2$}}}

\newcommand{\VPloopDt}[2]
{\put(0,0){\circle{1}} \put(0,0.4){\VectorR}
\put(0.05,-0.4){\Vector} \put(0,0.8){\makebox(0,0){$#1$}}
\put(0,-0.8){\makebox(0,0){$#2$}}}

\newcommand{\Loop}[2]
{\put(0,0){\oval(0.6,1.25)}\put(0.01,0.01){\oval(0.6,1.25)}\put(-0.01,-0.01){\oval(0.6,1.25)}
\put(0.3,0){\VectorUp}
\put(-0.3,0){\VectorDn}
\put(-0.65,0){\makebox(0,0){$#1$}}
\put(0.65,0){\makebox(0,0){$#2$}}}

\newcommand{\HFexch}[1]
{\put(0,0)\dashH
\qbezier(0,0.01)(0.625,0.515)(1.25,0.015)
\qbezier(0,-0.01)(0.625,0.485)(1.25,-0.015)
\put(0.625,0.26){\Vector}
\put(0.625,0.5){\makebox(0,0){$#1$}}
\put(0,0){\circle*{0.15}}
\put(1.25,0){\circle*{0.15}}}

\newcommand{\HFexcht}[1]
{\put(0,0)\dashH \qbezier(0,0.01)(0.625,0.515)(1.25,0.015)
\put(0.625,0.26){\Vector} \put(0.625,0.5){\makebox(0,0){$#1$}}
\put(0,0){\circle*{0.15}} \put(1.25,0){\circle*{0.15}}}

\newcommand{\dcircH}[2]
{\put(-0.5,0){\multiput(0.05,0)(0.25,0){8}{\line(1,0){0.15}}}
\put(-1,0){\makebox(0,0){$#1$}} \put(2,0){\makebox(0,0){$#2$}}
\put(0,0){\circl} \put(1,0){\circl}}

\newcommand{\dcirc}[2]
{\put(-0.5,0){\multiput(0.05,0)(0.25,0){12}{\line(1,0){0.15}}}
\put(-1,0){\makebox(0,0){$#1$}} \put(3,0){\makebox(0,0){$#2$}}
\put(0,0){\circl} \put(2,0){\circl}}

\newcommand{\dcircl}[2]
{\put(-0.5,0){\makebox(0,0){$#1$}} \put(2.5,0){\makebox(0,0){$#2$}}
\put(0,0){\circl} \put(2,0){\circl}}

\newcommand{\dcircOut}[5]
{\put(0,0){\dcirc{#1}{#2}} \put(0,0){\ellineP{#3}{2}{#4}{}{}{#5}}}

\newcommand{\dcircIn}[5]
{\put(0,0){\dcirc{#1}{#2}}
\put(0,-#3){\ellineP{#3}{2}{#4}{}{}{#5}}}
  
\newcommand{\dcircT}[2]
{\put(-0.5,0){\multiput(0.05,0)(0.25,0){16}{\line(1,0){0.15}}}
\put(-1,0){\makebox(0,0){$#1$}} \put(4,0){\makebox(0,0){$#2$}}
\put(0,0){\circl} \put(3,0){\circl}}

\newcommand{\dcircTIn}[4]
{\put(0,0){\dcircT{#1}{#2}}
\put(0,-1){\Elline{1}{0.5}{}{}}\put(3,-1){\Elline{1}{0.5}{}{}}
\put(-0.4,-0.5){\makebox(0,0){$#3$}}
\put(3.5,-0.5){\makebox(0,0){$#4$}}}

\newcommand{\dcircF}[2]
{\put(-0.5,0){\multiput(0.05,0)(0.25,0){20}{\line(1,0){0.15}}}
\put(-1,0){\makebox(0,0){$#1$}} \put(4,0){\makebox(0,0){$#2$}}
\put(0,0){\circl} \put(4,0){\circl}}

\newcommand{\dcircFIn}[4]
{\put(0,0){\dcircF{$#1$}{$#2$}}
\put(0,-1){\Elline{1}{0.5}{}{}}\put(4,-1){\Elline{1}{0.5}{}{}}
\put(-0.4,-0.5){\makebox(0,0){$#3$}}
\put(4.5,-0.5){\makebox(0,0){$#4$}}}

\newcommand{\dcircFem}[2]
{\put(-0.5,0){\multiput(0.05,0)(0.25,0){24}{\line(1,0){0.15}}}
\put(-1,0){\makebox(0,0){$#1$}} \put(5,0){\makebox(0,0){$#2$}}
\put(0,0){\circl} \put(5,0){\circl}}

\newcommand{\photonPP}
{\qbezier(0,0)(0.08333,0.125)(0.1666667,0)
\qbezier(0.1666667,0)(0.25,-0.125)(0.3333333,0)
\qbezier(0.3333333,0)(0.416667,0.125)(0.5,0)

\qbezier(0.5,0)(0.583333,-0.125)(0.666667,0)
\qbezier(0.666667,0)(0.75,0.125)(0.833333,0)
\qbezier(0.833333,0)(0.916667,-0.125)(1,0)}

\newcommand{\photonnh}
{\qbezier(0,0)(0.08333,0.125)(0.1666667,0)
\qbezier(0.1666667,0)(0.25,-0.125)(0.3333333,0)
\qbezier(0.3333333,0)(0.416667,0.125)(0.5,0)

\qbezier(0.5,0)(0.583333,-0.125)(0.666667,0)
\qbezier(0.666667,0)(0.75,0.125)(0.833333,0)
\qbezier(0.833333,0)(0.916667,-0.125)(1,0) }

\newcommand{\photonnH}
{\qbezier(0,0)(0.08333,0.125)(0.1666667,0)
\qbezier(0.1666667,0)(0.25,-0.125)(0.3333333,0)
\qbezier(0.3333333,0)(0.416667,0.125)(0.5,0)

\qbezier(0.5,0)(0.583333,-0.125)(0.666667,0)
\qbezier(0.666667,0)(0.75,0.125)(0.833333,0)
\qbezier(0.833333,0)(0.916667,-0.125)(1,0)

\qbezier(1,0)(1.083333,0.125)(1.166667,0)
\qbezier(1.166667,0)(1.25,-0.125)(1.333333,0)
\qbezier(1.333333,0)(1.416667,0.125)(1.5,0)}

\newcommand{\photonn}
{\qbezier(0,0)(0.08333,0.125)(0.1666667,0)
\qbezier(0.1666667,0)(0.25,-0.125)(0.3333333,0)
\qbezier(0.3333333,0)(0.416667,0.125)(0.5,0)

\qbezier(0.5,0)(0.583333,-0.125)(0.666667,0)
\qbezier(0.666667,0)(0.75,0.125)(0.833333,0)
\qbezier(0.833333,0)(0.916667,-0.125)(1,0)

\qbezier(1,0)(1.083333,0.125)(1.166667,0)
\qbezier(1.166667,0)(1.25,-0.125)(1.333333,0)
\qbezier(1.333333,0)(1.416667,0.125)(1.5,0)

\qbezier(1.5,0)(1.583333,-0.125)(1.666667,0)
\qbezier(1.666667,0)(1.75,0.125)(1.833333,0)
\qbezier(1.833333,0)(1.916667,-0.125)(2,0) }

\newcommand{\photonH}[3]
{\put(0,0){\photonPP}\put(1,0){\photonP}
\put(0.75,-0.15){\VectorR} \put(0.75,0.35){\makebox(0,0){$#1$}}
\put(0,0){\circl} \put(1.5,0){\circl}
\put(-0.5,0){\makebox(0,0){#2}} \put(2,0){\makebox(0,0){#3}}}

\newcommand{\photon}[3]
{\photonn\put(1,-0.05){\VectorR}
\put(1,0.35){\makebox(0,0.1){$#1$}} \put(0,0){\circl}
\put(2,0){\circl} \put(-0.5,0){\makebox(0,0){#2}}
\put(2.5,0){\makebox(0,0){#3}}}

\newcommand{\photonp}[3]
{\photonn \put(1,0.45){\makebox(0,0.1){$#1$}}
\put(-0.5,0){\makebox(0,0){#2}} \put(2.5,0){\makebox(0,0){#3}}}

\newcommand{\Photon}[3]
{
\linethickness{0.4mm} \put(0.,-0.00){\photonn{}{}{}}
 \put(0,0){\circl}\put(2,0){\circl}
}

\newcommand{\PPhoton}[3]
{\linethickness{0.7mm} \put(0.,-0.00){\photonn{}{}{}} }

\newcommand{\PPhotonH}[3]
{ \linethickness{0.7mm} \put(0.,-0.00){\photonnH{}{}{}} }

\newcommand{\PhotonH}[3]
{\put(0.,0.06){\photonnH{}{}{}}\put(0.,0.02){\photonnH{}{}{}}
\put(0.,-0.02){\photonnH{}{}{}}\put(0.,-0.06){\photonnH{}{}{}} }

\newcommand{\Photonh}[3]
{\put(0.,0.075){\photonnh{}{}{}}\put(0.,0.025){\photonnh{}{}{}}
\put(0.,-0.025){\photonnh{}{}{}}\put(0.,-0.075){\photonnh{}{}{}} }

\newcommand{\PhotonT}[3]
{\put(0.,0.){\photon{}{}{}} \put(1.,0){\photon{}{}{}}}

\newcommand{\photonT}[3]
{\put(0,0){\photonn{}{#2}{}}\put(1,0){\photonn{}{}{#3}}
\put(1.5,-0.05){\VectorR} \put(1.5,0.35){\makebox(0,0.1){$#1$}}
\put(0,0){\circl} \put(3,0){\circl}
\put(-0.5,0){\makebox(0,0){#2}} \put(3.5,0){\makebox(0,0){#3}}}

\newcommand{\photonF}[3]
{\put(0,0){\photonn{}{}{}}\put(2,0){\photonn{}{}{}}
\put(2,-0.05){\VectorR} \put(1,0.35){\makebox(0,0.1){$#1$}}
\put(0,0){\circl} \put(4,0){\circl}
\put(-0.5,0){\makebox(0,0){#2}} \put(4.5,0){\makebox(0,0){#3}}}

\newcommand{\photonHS}[4]
{\qbezier(0,0)(0.08333,0.125)(0.1666667,0)
\qbezier(0.1666667,0)(0.25,-0.125)(0.3333333,0)
\qbezier(0.3333333,0)(0.416667,0.125)(0.5,0)
\qbezier(0.5,0)(0.583333,-0.125)(0.666667,0)
\qbezier(0.666667,0)(0.75,0.125)(0.833333,0)
\qbezier(0.833333,0)(0.916667,-0.125)(1,0)
\put(0.5,-0.05){\VectorR} \put(0.5,0.35){\makebox(0,0){$#1$}}
\put(0,0){\circl}\put(1,0){\circl} \put(0,-0.5){\makebox(0,0){#2}}
\put(1,-0.5){\makebox(0,0){#3}}}

\newcommand{\photonNEn}[0]
{\qbezier(0,0)(0.22,-0.02)(0.2,0.2)
\qbezier(0.2,0.2)(0.18,0.42)(0.4,0.4)
\qbezier(0.4,0.4)(0.62,0.38)(0.6,0.6)
\qbezier(0.6,0.6)(0.58,0.82)(0.8,0.8)
\qbezier(0.8,0.8)(1.02,0.78)(1,1)
\qbezier(1,1)(0.98,1.22)(1.2,1.2)
\qbezier(1.2,1.2)(1.42,1.18)(1.4,1.4)
\qbezier(1.4,1.4)(1.38,1.62)(1.6,1.6)
\qbezier(1.6,1.6)(1.82,1.58)(1.8,1.8)
\qbezier(1.8,1.8)(1.78,2.02)(2,2) }

\newcommand{\photonNEu}[0]
{\qbezier(0,0)(-0.02,0.22)(0.2,0.2)
\qbezier(0.2,0.2)(0.42,0.18)(0.4,0.4)
\qbezier(0.4,0.4)(0.38,0.62)(0.6,0.6)
\qbezier(0.6,0.6)(0.82,0.58)(0.8,0.8)
\qbezier(0.8,0.8)(0.78,1.02)(1,1)
\qbezier(1,1)(1.22,0.98)(1.2,1.2)
\qbezier(1.2,1.2)(1.18,1.42)(1.4,1.4)
\qbezier(1.4,1.4)(1.62,1.38)(1.6,1.6)
\qbezier(1.6,1.6)(1.58,1.82)(1.8,1.8)
\qbezier(1.8,1.8)(2.02,1.78)(2,2) }

\newcommand{\photonNE}[3]
{\photonNEn \put(1,1){\makebox(0.05,-0.2){\VectorUp}}
\put(0,0){\circl} \put(2,2){\circl}
\put(0.8,0.6){\makebox(0,0){$#1$}}
\put(-0.35,-1){\makebox(0,2){$#2$}}
\put(2.35,1){\makebox(0,2){$#3$}}}

\newcommand{\photonNNE}[3]
{\qbezier(0,0)   (0.28,-0.02)(0.2,0.3)
\qbezier(0.2,0.3)(0.12,0.52)(0.4,0.6)
\qbezier(0.4,0.6)(0.68,0.58)(0.6,0.9)
\qbezier(0.6,0.9)(0.52,1.12)(0.8,1.2)
\qbezier(0.8,1.2)(1.08,1.18)(1,1.5) \qbezier(1,1.5)
(0.92,1.72)(1.2,1.8) \qbezier(1.2,1.8)(1.48,1.86)(1.4,2.1)
\qbezier(1.4,2.1)(1.365,2.24)(1.6,2.4)
\qbezier(1.6,2.4)(1.835,2.46)(1.8,2.7)
\qbezier(1.8,2.7)(1.765,2.84)(2,3)
\put(0.6,0.8){\makebox(0,0){\VectorUp}} \put(0,0){\circle*{0.15}}
\put(2,3){\circle*{0.15}} \put(1,0.8){\makebox(0,0){$#1$}}
\put(-0.35,0){\makebox(0,0){$#2$}}
\put(2.35,3){\makebox(0,0){$#3$}}}

\newcommand{\photonENE}[3]
{\qbezier(0,0)(0.17,-0.04)(0.2,0.1)
\qbezier(0.2,0.1)(0.23,0.32)(0.4,0.2)
\qbezier(0.4,0.2)(0.57,0.16)(0.6,0.3)
\qbezier(0.6,0.3)(0.63,0.52)(0.8,0.4)
\qbezier(0.8,0.4)(0.97,0.36)(1,0.5)
\qbezier(1,0.5)(1.03,0.72)(1.2,0.6)
\qbezier(1.2,0.6)(1.37,0.56)(1.4,0.7)
\qbezier(1.4,0.7)(1.43,0.92)(1.6,0.8)
\qbezier(1.6,0.8)(1.77,0.76)(1.8,0.9)
\qbezier(1.8,0.9)(1.83,1.12)(2,1)
\put(1,0.8){\makebox(0,0){\VectorR}} \put(0,0){\circle*{0.15}}
\put(2,1){\circle*{0.15}} \put(1,0.85){\makebox(0,-0){$#1$}}
\put(-0.35,-1){\makebox(0,2){$#2$}}
\put(2.35,0){\makebox(0,2){$#3$}}}

\newcommand{\photonENEh}[3]
{\qbezier(0,0)(0.17,-0.04)(0.2,0.1)
\qbezier(0.2,0.1)(0.23,0.32)(0.4,0.2)
\qbezier(0.4,0.2)(0.57,0.16)(0.6,0.3)
\qbezier(0.6,0.3)(0.63,0.52)(0.8,0.4)
\qbezier(0.8,0.4)(0.97,0.36)(1,0.5)
\put(1,0.85){\makebox(0,-0){$#1$}}
\put(-0.35,-1){\makebox(0,2){$#2$}}
\put(2.35,0){\makebox(0,2){$#3$}}}

\newcommand{\photonNWn}[0]
{\qbezier(0,0)(-0.22,-0.02)(-0.2,0.2)
\qbezier(-0.2,0.2)(-0.18,0.42)(-0.4,0.4)
\qbezier(-0.4,0.4)(-0.62,0.38)(-0.6,0.6)
\qbezier(-0.6,0.6)(-0.58,0.82)(-0.8,0.8)
\qbezier(-0.8,0.8)(-1.02,0.78)(-1,1) \qbezier(-1,1)
(-0.98,1.22)(-1.2,1.2) \qbezier(-1.2,1.2)(-1.42,1.18)(-1.4,1.4)
\qbezier(-1.4,1.4)(-1.38,1.62)(-1.6,1.6)
\qbezier(-1.6,1.6)(-1.82,1.58)(-1.8,1.8)
\qbezier(-1.8,1.8)(-1.78,2.02)(-2,2)}

\newcommand{\photonNWu}[0]
{\qbezier(0,0)(0.02,-0.22)(-0.2,0.2)
\qbezier(-0.2,0.2)(-0.42,0.18)(-0.4,0.4)
\qbezier(-0.4,0.4)(-0.38,0.62)(-0.6,0.6)
\qbezier(-0.6,0.6)(-0.82,0.58)(-0.8,0.8)
\qbezier(-0.8,0.8)(-0.78,1.02)(-1,1)
\qbezier(-1,1)(-1.22,0.98)(-1.2,1.2)
\qbezier(-1.2,1.2)(-1.18,1.42)(-1.4,1.4)
\qbezier(-1.4,1.4)(-1.62,1.38)(-1.6,1.6)
\qbezier(-1.6,1.6)(-1.58,1.8)(-1.8,1.8)
\qbezier(-1.8,1.8)(-2.02,1.78)(-2,2)}

\newcommand{\photonNW}[3]
{\put(0,0)\photonNWn \put(-1,1){\makebox(0,-0.2){\VectorUp}}
\put(0,0){\circl} \put(-2,2){\circl}
\put(-1.1,1.4){\makebox(0,0){$#1$}}
\put(-2.35,2){\makebox(0,0){$#2$}}
\put(0.35,0){\makebox(0,0){$#3$}}}

\newcommand{\photonTNE}[3]
{\photonNEn \put(1,1)\photonNEu
\put(1.5,1.5){\makebox(0.05,-0.2){\VectorUp}} \put(0,0){\circl}
\put(3,3){\circl} \put(1,1){\makebox(-0.4,1){$#1$}}
\put(-0.35,0){\makebox(0,0){$#2$}}
\put(3.35,3.2){\makebox(0,0){$#3$}}}

\newcommand{\photonTNW}[3]
{\photonNWn \put(-1,1)\photonNWu
\put(-1.5,1.5){\makebox(0.05,-0.2){\VectorUp}} \put(0,0){\circl}
\put(-3,3){\circl} \put(1,1){\makebox(-0.4,1){$#1$}}
\put(-0.35,-1){\makebox(0,2){$#2$}}
\put(3.35,1){\makebox(0,2){$#3$}}}

\newcommand{\photonFNW}[3]
{\photonNWn \put(-2,2)\photonNWu
\put(-2,2){\makebox(0.05,-0.2){\VectorUp}} \put(0,0){\circl}
\put(-4,4){\circl} \put(1,1){\makebox(-0.4,1){$#1$}}
\put(-0.35,-1){\makebox(0,2){$#2$}}
\put(4.35,1){\makebox(0,2){$#3$}}}

\newcommand{\photonSEst}[3]
{\qbezier(0,0)(-0.22,-0.02)(-0.2,0.2)
\qbezier(-0.2,0.2)(-0.18,0.42)(-0.4,0.4)
\qbezier(-0.4,0.4)(-0.62,0.38)(-0.6,0.6)
\qbezier(-0.6,0.6)(-0.58,0.82)(-0.8,0.8)
\qbezier(-0.8,0.8)(-1.02,0.78)(-1,1) \qbezier(-1,1)
(-0.98,1.22)(-1.2,1.2) \qbezier(-1.2,1.2)(-1.42,1.18)(-1.4,1.4)
\qbezier(-1.4,1.4)(-1.38,1.62)(-1.6,1.6)
\qbezier(-1.6,1.6)(-1.82,1.58)(-1.8,1.8)
\qbezier(-1.8,1.8)(-1.78,2.02)(-2,2)
\put(-1,1){\makebox(0,0){\VectorDn}} \put(0,0){\circle*{0.15}}
\put(-2,2){\circle*{0.15}} \put(-1,1){\makebox(0.4,0.7){$#1$}}
\put(-2.35,2){\makebox(0,0){$#2$}}
\put(0.35,0){\makebox(0,0){$#3$}}}

\newcommand{\photonWNW}[3]
{\qbezier(0,0)(-0.17,-0.04)(-0.2,0.1)
\qbezier(-0.2,0.1)(-0.23,0.32)(-0.4,0.2)
\qbezier(-0.4,0.2)(-0.57,0.16)(-0.6,0.3)
\qbezier(-0.6,0.3)(-0.63,0.52)(-0.8,0.4)
\qbezier(-0.8,0.4)(-0.97,0.36)(-1,0.5)
\qbezier(-1,0.5)(-1.03,0.72)(-1.2,0.6)
\qbezier(-1.2,0.6)(-1.37,0.56)(-1.4,0.7)
\qbezier(-1.4,0.7)(-1.43,0.92)(-1.6,0.8)
\qbezier(-1.6,0.8)(-1.77,0.76)(-1.8,0.9)
\qbezier(-1.8,0.9)(-1.83,1.12)(-2,1)
\put(-1.1,0.75){\makebox(0,0){$\;$\VectorR}}
\put(0,0){\circle*{0.15}} \put(-2,1){\circle*{0.15}}
\put(-1,0.5){\makebox(0.6,0.4){$#1$}}
\put(0.35,-1){\makebox(0,2){$#3$}}
\put(-2.35,0){\makebox(0,2){$#2$}}}

\newcommand{\Crossphotons}[6]
{\put(0,0){\photonNE{#5}{}{}}
\put(2,0){\photonNW{#6}{}{}}
\put(-0.35,0){\makebox(0,0){$#1$}}
\put(2.35,2){\makebox(0,0){$#2$}}
\put(2.35,0){\makebox(0,0){$#3$}}
\put(-0.35,2){\makebox(0,0){$#4$}}
}

\newcommand{\photonNe}[3]
{\qbezier(0,0)(0.22,-0.02)(0.2,0.2)
\qbezier(0.2,0.2)(0.18,0.42)(0.4,0.4)
\qbezier(0.4,0.4)(0.62,0.38)(0.6,0.6)
\qbezier(0.6,0.6)(0.58,0.82)(0.8,0.8)
\qbezier(0.8,0.8)(1.02,0.78)(1,1)
\qbezier(1,1)(0.98,1.22)(1.2,1.2)
\put(0.75,0.75){\makebox(-0.6,0.4){$#1$}}
\put(-0.35,0){\makebox(0,0){$#2$}}
\put(1.85,1.5){\makebox(0,0){$#3$}}}

\newcommand{\PhotonNe}[3]
{\put(0,0){\photonNe{}{}{}}\put(0.05,-0.05){\photonNe{}{}{}}
\put(-0.05,0.05){\photonNe{}{}{}}
\put(0.75,0.75){\makebox(-0.6,0.4){$#1$}}
\put(-0.35,0){\makebox(0,0){$#2$}}
\put(1.85,1.5){\makebox(0,0){$#3$}}}

\newcommand{\photonNw}[3]
{\qbezier(0,0)(0.02,0.22)(-0.2,0.2)
\qbezier(-0.2,0.2)(-0.42,0.18)(-0.4,0.4)
\qbezier(-0.4,0.4)(-0.38,0.62)(-0.6,0.6)
\qbezier(-0.6,0.6)(-0.82,0.58)(-0.8,0.8)
\qbezier(-0.8,0.8)(-0.78,1.02)(-1,1) \qbezier(-1,1)
(-1.22,0.98)(-1.2,1.2) \qbezier(-1.2,1.2)(-1.18,1.42)(-1.4,1.4)
\qbezier(-1.4,1.4)(-1.5,1.38)(-1.5,1.5)
\put(0,0){\circle*{0.20}}
\put(-0.7,0.7){\makebox(1,0.3){$#1$}}
\put(0.35,0){\makebox(0,0){$#3$}}
\put(-1.85,1.5){\makebox(0,0){$#2$}} }

\newcommand{\elstat}[3]
{\multiput(0.06,0)(0.35,0){6}{\line(1,0){0.15}}
\put(1,0.35){\makebox(0,0){$#1$}} \put(-0.35,0){\makebox(0,0){#2}}
\put(0,0){\circl}\put(2,0){\circl}
 \put(2.35,0){\makebox(0,0){#3}}}

\newcommand{\PairOut}[6]
 {\Pair{#1}{#2}{#3}\put(0,0){\elline{#4}{#5}{}}
 \put(#1,0){\elline{#4}{}{#6}}}

\newcommand{\Pair}[3]
{{\linethickness{1mm} \put(0,0){\line(1,0){#1}}}
 \put(-0.3,0){\makebox(0,0){$#2$}}
\put(0,0){\circl}\put(#1,0){\circl}
 \put(#1,0){\makebox(0.6,0){$#3$}}}

\newcommand{\PPair}[3]
{\linethickness{2mm} \put(0,0){\line(1,0){#1}}
 \put(-0.3,0){\makebox(0,0){$#2$}}
 \put(#1,0){\makebox(0.6,0){$#3$}}}

\newcommand{\OmQED}[3]
{\linethickness{2mm} \put(0,0){\line(1,0){#1}}
 \put(-0.3,0){\makebox(0,0){$#2$}}
 \put(#1,0){\makebox(0.6,0){$#3$}}}

\newcommand{\OmQEDP}[3]
{\linethickness{2mm} \put(0,0){\line(1,0){#1}}
 \put(-0.3,0){\makebox(0,0){$#2$}}
 \put(#1,0){\makebox(0.6,0){$#3$}}}

\newcommand{\PPairP}
{\linethickness{2mm} \put(0,0){\line(1,0){2}}
}

\newcommand{\PPairH}
{\linethickness{1.5mm} \put(0,0){\line(1,0){1.5}}
}

\newcommand{\PPairh}
{\linethickness{1.5mm} \put(0,0){\line(1,0){1}}
}

\newcommand{\UQED}
{\put(0,-0.125){\Elstat{}{}{}}\put(0,0.15){\Elstat{}{}{}}
\put(0.,0.){\Photon{}{}{}}}

\newcommand{\UQEDH}
{\put(0,-0.125){\ElstatH{}{}{}}\put(0,0.15){\ElstatH{}{}{}}
\put(0.,0.){\PhotonH{}{}{}}}

\newcommand{\UQEDh}
{\put(0,-0.125){\Elstath{}{}{}}\put(0,0.15){\Elstath{}{}{}}
\put(0.,0.){\Photonh{}{}{}}}

\newcommand{\Usp}
{{\linethickness{0.3mm}\put(0,-0.125){\line(1,0){2}}\put(0,0.125){\line(1,0){2}}}
\put(0.,0.){\photonn{}{}{}}}

\newcommand{\UspH}
{{\linethickness{0.3mm}\put(0,-0.125){\line(1,0){1.5}}\put(0,0.125){\line(1,0){1.5}}}
\put(0.,0.){\photonnH{}{}{}}}

\newcommand{\Usph}
{{\linethickness{0.3mm}\put(0,-0.125){\line(1,0){1}}\put(0,0.125){\line(1,0){1}}}
\put(0.,0.){\photonnh{}{}{}}}

\newcommand{\PairQEDT}
{\put(0,0){\PairQED}\put(1.,0){\PairQED}}

\newcommand{\PairQEDH}
{\linethickness{0.2mm}\put(0,-0.125){\line(1,0){1.5}}\put(0,0.125){\line(1,0){1.5}}
\put(0.,0.){\PhotonH{}{}{}}}

\newcommand{\PairQEDh}
{\put(0,0){\Pair{1}{}{}}\put(0.,0.1){\photonnh{}{}{}}\put(0.,-0.1){\photonnh{}{}{}}
}

\newcommand{\PairIn}[6]
 {\Pair{#1}{#2}{#3}\put(0,-#4){\elline{#4}{#5}{}}
 \put(#1,-#4){\elline{#4}{}{#6}}}

\newcommand{\CoulIn}[5]
 {\Elstat{}{#1}{#2}\put(0,-#3){\elline{#3}{#4}{}}
 \put(2,-#3){\elline{#3}{}{#5}}}

\newcommand{\CoulInT}[5]
 {\ElstatT{}{#1}{#2}\put(0,-#3){\elline{#3}{#4}{}}
 \put(3,-#3){\elline{#3}{}{#5}}}

\newcommand{\CoulInF}[5]
 {\ElstatF{}{#1}{#2}\put(0,-#3){\elline{#3}{#4}{}}
 \put(4,-#3){\elline{#3}{}{#5}}}

 \newcommand{\CoulInh}[5]
 {\Elstath{}{#1}{#2}\put(0,-#3){\elline{#3}{#4}{}}
 \put(1,-#3){\elline{#3}{}{#5}}}

\newcommand{\elstatn}[3]
{\multiput(0.06,0)(0.25,0){8}{\line(1,0){0.15}}
\put(1,0.35){\makebox(0,0){$#1$}} \put(-0.35,0){\makebox(0,0){#2}}
 \put(2.35,0){\makebox(0,0){#3}}}

\newcommand{\Elstat}[3]
{\linethickness{0.3mm}\multiput(0.06,0)(0.35,0){6}{\line(1,0){0.15}}
\put(1,0.35){\makebox(0,0){$#1$}}
 \put(0,0){\circl}\put(2,0){\circl}
 \put(-0.35,0){\makebox(0,0){#2}}
\put(2.35,0){\makebox(0,0){#3}}}

\newcommand{\Elstatn}[3]
{\linethickness{0.3mm}\multiput(0.06,0)(0.35,0){6}{\line(1,0){0.15}}
\put(1,0.35){\makebox(0,0){$#1$}}}

\newcommand{\elstatT}[3]
{\linethickness{0.4mm}\multiput(0.06,0)(0.333,0){9}{\line(1,0){0.15}}
\put(1,0.35){\makebox(0,0){$#1$}} \put(0,0){\circl}
\put(3,0){\circl} \put(-0.35,0){\makebox(0,0){#2}}
\put(3.35,0){\makebox(0,0){#3}}}

\newcommand{\ElstatT}[3]
{\linethickness{0.4mm}\multiput(0.06,0)(0.333,0){9}{\line(1,0){0.15}}
\put(1,0.35){\makebox(0,0){$#1$}} \put(0,0){\circl}
\put(3,0){\circl} \put(-0.35,0){\makebox(0,0){#2}}
\put(3.35,0){\makebox(0,0){#3}}}

\newcommand{\ElstatF}[3]
{\linethickness{0.4mm}\multiput(0.06,0)(0.333,0){12}{\line(1,0){0.15}}
\put(1,0.35){\makebox(0,0){$#1$}} \put(0,0){\circl}
\put(4,0){\circl} \put(-0.35,0){\makebox(0,0){#2}}
\put(4.35,0){\makebox(0,0){#3}}}

\newcommand{\elstatF}[3]
{\multiput(0.06,0)(0.333,0){12}{\line(1,0){0.15}}
\put(1,0.35){\makebox(0,0){$#1$}} \put(0,0){\circl}
\put(4,0){\circl} \put(-0.35,0){\makebox(0,0){#2}}
\put(4.35,0){\makebox(0,0){#3}}}

\newcommand{\Breit}[3]
{\multiput(0.0,0)(0.4,0){6}{\circle*{0.2}}
\put(1,0.35){\makebox(0,0){$#1$}}
  \put(-0.35,0){\makebox(0,0){#2}}
\put(2.35,0){\makebox(0,0){#3}}}

\newcommand{\Melstat}[1]
{\multiput(0,0)(0,0.2){#1}{\Elstat{}{}{}}}

\newcommand{\MelstatT}[1]
{\multiput(0,0)(0,0.2){#1}{\elstatT{}{}{}}}

\newcommand{\Multiline}[3]
{\linethickness{0.2mm} \put(0,-0.1){\line(1,0){#1}}
 \put(0,0){\line(1,0){#1}}\put(0,0.1){\line(1,0){#1}}
 \put(-0.35,0){\makebox(0,0){#2}}
\put(2.35,0){\makebox(0,0){#3}}}

\newcommand{\elstatH}[3]
{\multiput(0.06,0)(0.25,0){6}{\line(1,0){0.15}}
\put(0.75,0.25){\makebox(0,0){$#1$}} \put(0,0){\circle*{0.1}}
\put(1.5,0){\circle*{0.1}}
\put(-0.35,0){\makebox(0,0){#2}}
\put(2.35,0){\makebox(0,0){#3}}}

\newcommand{\ElstatH}[3]
{\linethickness{0.4mm}\multiput(0.06,0)(0.25,0){6}{\line(1,0){0.15}}
\put(0.75,0.25){\makebox(0,0){$#1$}} \put(0,0){\circl}
\put(1.5,0){\circl}
\put(-0.35,0){\makebox(0,0){#2}} \put(2.35,0){\makebox(0,0){#3}}}

\newcommand{\elstath}[3]
{\multiput(0.06,0)(0.25,0){4}{\line(1,0){0.15}}
\put(0.75,0.25){\makebox(0,0){$#1$}} 
\put(-0.35,0){\makebox(0,0){#2}} \put(2.35,0){\makebox(0,0){#3}}}

\newcommand{\Elstath}[3]
{\linethickness{0.4mm}
\multiput(0.06,0)(0.25,0){4}{\line(1,0){0.15}}
\put(0.75,0.25){\makebox(0,0){$#1$}} \put(0,0){\circle*{0.1}}
\put(1.5,0){\circle*{0.1}}
\put(-0.35,0){\makebox(0,0){#2}} \put(2.35,0){\makebox(0,0){#3}}}

\newcommand{\BreitH}[3]
{\multiput(0.15,0)(0.3,0){5}{\circle*{0.1}}
\put(0.75,0.25){\makebox(0,0){$#1$}} 
\put(-0.35,0){\makebox(0,0){#2}}
\put(2.35,0){\makebox(0,0){#3}}}

\newcommand{\RetBreitDH}[3]
{\small\multiput(0.15,0.3)(0.3,-0.15){5}{\circle*{0.1}}
\put(0.75,0.25){\makebox(0,0){$#1$}}
\put(-0.35,0){\makebox(0,0){#2}} \put(2.35,0){\makebox(0,0){#3}}}

\newcommand{\RetBreitH}[3]
{\small\multiput(0.15,-0.3)(0.3,0.15){5}{\circle*{0.1}}
\put(0.75,0.25){\makebox(0,0){$#1$}}
\put(-0.35,0){\makebox(0,0){#2}} \put(2.35,0){\makebox(0,0){#3}}}

\newcommand{\elsta}[3]
{\multiput(0.06,0)(0.25,0){4}{\line(1,0){0.15}}
\put(1,0.35){\makebox(0,0){$#1$}}
\put(0,0){\circle*{0.1}}
\put(1,0){\circle*{0.1}}
\put(-0.35,0){\makebox(0,0){#2}}
\put(1.35,0){\makebox(0,0){#3}}}

\newcommand{\elstatNO}[3]
{\multiput(0.06,0.08)(0.01,0.01){14}{\tiny.}
\multiput(0.30,0.32)(0.01,0.01){14}{\tiny.}
\multiput(0.55,0.57)(0.01,0.01){14}{\tiny.}
\multiput(0.79,0.81)(0.01,0.01){14}{\tiny.}
\multiput(1.03,1.05)(0.01,0.01){14}{\tiny.}
\multiput(1.27,1.29)(0.01,0.01){14}{\tiny.}
\multiput(1.51,1.53)(0.01,0.01){14}{\tiny.}
\multiput(1.75,1.77)(0.01,0.01){14}{\tiny.}
\put(0.95,0.75){\makebox(0,0){\VectorUr}}
\put(0,0){\circle*{0.15}}
\put(2,2){\circle*{0.15}}
\put(0.75,1.25){\makebox(0,0){$#1$}}
\put(-0.5,0){\makebox(0,0){$#2$}}
\put(2.5,2){\makebox(0,0){$#3$}}
}

\newcommand{\elstatNW}[3]
{\multiput(-0.05,0.05)(-0.015,0.015){10}{\circle*{0.02}}
\multiput(-0.3,0.3)(-0.015,0.015){10}{\circle*{0.02}}
\multiput(-0.55,0.55)(-0.015,0.015){10}{\circle*{0.02}}
\multiput(-0.8,0.8)(-0.015,0.015){10}{\circle*{0.03}}
\multiput(-1.05,1.05)(-0.015,0.015){10}{\circle*{0.03}}
\multiput(-1.3,1.3)(-0.015,0.015){10}{\circle*{0.03}}
\multiput(-1.55,1.55)(-0.015,0.015){10}{\circle*{0.03}}
\multiput(-1.8,1.8)(-0.015,0.015){10}{\circle*{0.03}}
\put(-0.9,0.9){\makebox(0,0){\VectorUl}}
\put(0,0){\circle*{0.215}}
\put(-2,2){\circle*{0.2153}}
\put(-0.9,0.9){\makebox(0,0){\VectorUl}}
\put(0,0){\circle*{0.215}}
\put(-2,2){\circle*{0.215}}
\put(-0.5,1){\makebox(0,0){$#1$}}
\put(0,-0.5){\makebox(0,0){$#2$}}
\put(-2,2.5){\makebox(0,0){$#3$}}
}

\newcommand{\photonSE}[6]
{\put(0,0){\photonHS{#3}{#4}{}{}} \put(1.5,0){\VPloopD{#1}{#2}}
\put(2,0){\photonHS{#5}{#6}{}{}}}

\newcommand{\photonSEt}[5]
{\put(0,0){\photonHS{#3}{#4}{}{}}
\put(1.5,0){\VPloopDt{#1}{#2}}
\put(2,0){\photonHS{#5}{}{}{}}}

\newcommand{\ElSE}[3]
{\qbezier(0,-1)(.2025,-1.1489)(0.3420,-0.9397)
\qbezier(0.3420,-0.9397)(0.4167,-0.7217)(0.6428,-0.766)
\qbezier(0.6428,-0.766)(0.8937,-0.7499)(0.866,-0.5)
\qbezier(0.866,-0.5)(0.7831,-0.2850)(0.9848,-0.1736)
\qbezier(0.9848,-0.1736)(1.1667,0)(0.9848,0.1736)
\qbezier(0.9848,0.1736)(0.7831,0.2850)(0.866,0.5)
\qbezier(0.866,0.5)(0.8937,0.7499)(0.6428,0.766)
\qbezier(0.6428,0.766)(0.4167,0.7217)(0.3420,0.9397)
\qbezier(0.3420,0.9397)(.2025,1.1489)(0,1) \put(1,0.02){\VectorUp}
\put(0,1){\circl} \put(0,-1){\circl}
\put(1.45,0){\makebox(0,0){$#1$}} \put(-0.35,-1){\makebox(0,0){#2}}
\put(-0.35,1){\makebox(0,0){#3}}}

\newcommand{\ElSEN}[3]
{\qbezier(0,-1)(.1718,-1.0420)(0.1863,-0.8399)
\qbezier(0.1863,-0.8399)(0.1704,-0.6495)(0.3357,-0.6346)
\qbezier(0.3357,-0.6346)(0.5055,-0.5831)(0.4399,-0.3950)
\qbezier(0.4399,-0.3950)(0.3517,-0.2330)(0.4935,-0.1341)
\qbezier(0.4935,-0.1341)(0.6252,0)(0.4935,0.1341)
\qbezier(0.4935,0.1341)(0.3517,0.2330)(0.4399,0.3950)
\qbezier(0.4399,0.3950)(0.5055,0.5831)(0.3357,0.6346)
\qbezier(0.3357,0.6346)(0.1704,0.6495)(0.1863,0.8399)
\qbezier(0.1863,0.8399)(0.1718,1.0420)(0,1)
\put(0.5,0.02){\VectorUp} \put(0,1){\circle*{0.15}}
\put(0,-1){\circle*{0.15}} \put(0.85,0){\makebox(0,0){$#1$}}
\put(-0.35,-1){\makebox(0,0){#2}}
\put(-0.35,1){\makebox(0,0){#3}}}

\newcommand{\ElSENR}[3]
{\qbezier(-0.4472,-0.8944)(-0.3123,-1.0088)(-0.209,-0.8346)
\qbezier(-0.2090,-0.8346)(-0.138,-0.6572)(0.0164,-0.7177)
\qbezier(0.01664,-0.7177)(0.1914,-0.7476)(0.2169,-0.55)
\qbezier(0.2169,-0.55)(0.2103,-0.3657)(0.3815,-0.3406)
\qbezier(0.3815,-0.3406)(0.5593,-0.2796)(0.5014,-0.1079)
\qbezier(0.5014,-0.1008)(0.4188,0.0512)(0.5701,0.1565)
\qbezier(0.5701,0.1565)(0.7129,0.2954)(0.584,0.4175)
\qbezier(0.584,0.4175)(0.4429,0.5047)(0.5423,0.6679)
\qbezier(0.5423,0.6679)(0.6617,0.8552)(0.4472,0.8944)
\put(0.5,-0.1){\VectorUp} \put(0.475,0.95){\circle*{0.15}}
\put(-0.475,-0.95){\circle*{0.15}}
\put(0.9,0){\makebox(0,0){$#1$}} \put(-0.85,-1){\makebox(0,0){#2}}
\put(0.15,1){\makebox(0,0){#3}}}

\newcommand{\ElSENL}[3]
{\qbezier(-0.4472,0.8944)(-0.3123,1.0088)(-0.209,0.8346)
\qbezier(-0.2090,0.8346)(-0.138,0.6572)(0.0164,0.7177)
\qbezier(0.01664,0.7177)(0.1914,0.7476)(0.2169,0.55)
\qbezier(0.2169,0.55)(0.2103,0.3657)(0.3815,0.3406)
\qbezier(0.3815,0.3406)(0.5593,0.2796)(0.5014,0.1079)
\qbezier(0.5014,0.1008)(0.4188,-0.0512)(0.5701,-0.1565)
\qbezier(0.5701,-0.1565)(0.7129,-0.2954)(0.584,-0.4175)
\qbezier(0.584,-0.4175)(0.4429,-0.5047)(0.5423,-0.6679)
\qbezier(0.5423,-0.6679)(0.6617,-0.8552)(0.4472,-0.8944)
\put(0.5,0.1){\VectorUp} \put(-0.475,0.95){\circle*{0.15}}
\put(0.475,-0.95){\circle*{0.15}}
\put(-0.9,0){\makebox(0,0){$#1$}} \put(0.85,-1){\makebox(0,0){#2}}
\put(-0.85,1){\makebox(0,0){#3}}}

\newcommand{\ElSEL}[3]
{\qbezier(0,-1)(-.2025,-1.1489)(-0.3420,-0.9397)
\qbezier(-0.3420,-0.9397)(-0.4167,-0.7217)(-0.6428,-0.766)
\qbezier(-0.6428,-0.766)(-0.8937,-0.7499)(-0.866,-0.5)
\qbezier(-0.866,-0.5)(-0.7831,-0.2850)(-0.9848,-0.1736)
\qbezier(-0.9848,-0.1736)(-1.1667,0)(-0.9848,0.1736)
\qbezier(-0.9848,0.1736)(-0.7831,0.2850)(-0.866,0.5)
\qbezier(-0.866,0.5)(-0.8937,0.7499)(-0.6428,0.766)
\qbezier(-0.6428,0.766)(-0.4167,0.7217)(-0.3420,0.9397)
\qbezier(-0.3420,0.9397)(-.2025,1.1489)(0,1)
\put(-1,0.02){\VectorUp} 
\put(-1.45,0){\makebox(0,0){$#1$}}
\put(0.35,-1){\makebox(0,0){#2}} \put(0.35,1){\makebox(0,0){#3}}}

\newcommand{\SEpolt}[5]
{\qbezier(0,-1.5)(.2025,-1.6489)(0.3420,-1.4397)
\qbezier(0.3420,-1.4397)(0.4167,-1.2217)(0.6428,-1.266)
\qbezier(0.6428,-1.266)(0.8937,-1.2499)(0.866,-1)
\qbezier(0.866,-1)(0.7831,-0.7850)(0.9848,-0.6736)
\qbezier(1,-0.5)(1.1,-0.5)(0.9848,-0.6736)
\qbezier(1,0.5)(1.1,0.5)(0.9848,0.6736)
\qbezier(0.9848,0.6736)(0.7831,0.7850)(0.866,1)
\qbezier(0.866,1)(0.8937,1.2499)(0.6428,1.266)
\qbezier(0.6428,1.266)(0.4167,1.2217)(0.3420,1.4397)
\qbezier(0.3420,1.4397)(.2025,1.6489)(0,1.5)
\put(1,0){\VPloopLR{#1}{#2}} \put(0.87,-1){\VectorUp}
\put(0.67,1.23){\Vector} \put(1.3,-1){\makebox(0,0){#3}}
\put(1,0.5){\circle*{0.15}} \put(1,-0.5){\circle*{0.15}}
\put(0,1.5){\circle*{0.15}} \put(0,-1.5){\circle*{0.15}}
\put(-0.35,-1.5){\makebox(0,0){#4}}
\put(-0.35,1.5){\makebox(0,0){#5}}}

\newcommand{\SEpoltNA}[5]
{\qbezier(0,-1.5)(.2025,-1.6489)(0.3420,-1.4397)
\qbezier(0.3420,-1.4397)(0.4167,-1.2217)(0.6428,-1.266)
\qbezier(0.6428,-1.266)(0.8937,-1.2499)(0.866,-1)
\qbezier(0.866,-1)(0.7831,-0.7850)(0.9848,-0.6736)
\qbezier(1,-0.5)(1.1,-0.5)(0.9848,-0.6736)
\qbezier(1,0.5)(1.1,0.5)(0.9848,0.6736)
\qbezier(0.9848,0.6736)(0.7831,0.7850)(0.866,1)
\qbezier(0.866,1)(0.8937,1.2499)(0.6428,1.266)
\qbezier(0.6428,1.266)(0.4167,1.2217)(0.3420,1.4397)
\qbezier(0.3420,1.4397)(.2025,1.6489)(0,1.5)
\put(1,0){\circle{1}}
\put(0.87,-1){\VectorUp}
\put(0.67,1.23){\Vector}
\put(1.3,-1){\makebox(0,0){#3}}
\put(1,0.5){\circle*{0.15}}
\put(1,-0.5){\circle*{0.15}}
\put(0,1.5){\circle*{0.15}}
\put(0,-1.5){\circle*{0.15}}
\put(-0.35,-1.5){\makebox(0,0){#4}}
\put(-0.35,1.5){\makebox(0,0){#5}}}

\newcommand{\SEpol}[5]
{\qbezier(0,-1.5)(.2025,-1.6489)(0.3420,-1.4397)
\qbezier(0.3420,-1.4397)(0.4167,-1.2217)(0.6428,-1.266)
\qbezier(0.6428,-1.266)(0.8937,-1.2499)(0.866,-1)
\qbezier(0.866,-1)(0.7831,-0.7850)(0.9848,-0.6736)
\qbezier(1,-0.5)(1.1,-0.5)(0.9848,-0.6736)
\qbezier(1,0.5)(1.1,0.5)(0.9848,0.6736)
\qbezier(0.9848,0.6736)(0.7831,0.7850)(0.866,1)
\qbezier(0.866,1)(0.8937,1.2499)(0.6428,1.266)
\qbezier(0.6428,1.266)(0.4167,1.2217)(0.3420,1.4397)
\qbezier(0.3420,1.4397)(.2025,1.6489)(0,1.5)
\put(1,0){\VPloopLR{#1}{#2}}
\put(0.87,-1){\VectorUp}
\put(0.67,1.23){\Vector}
\put(1.3,-1){\makebox(0,0){#3}}
\put(1,0.5){\circle*{0.15}}
\put(1,-0.5){\circle*{0.15}}
\put(0,1.5){\circle*{0.15}}
\put(0,-1.5){\circle*{0.15}}
\put(-0.35,-1.5){\makebox(0,0){#4}}
\put(-0.35,1.5){\makebox(0,0){#5}}}

%% file: FigGreen.tex

\newcommand{\SEboxT}[2]
{\put(0,0){\LineH{#1}} \put(0,#2){\LineH{#1}}
\put(0,0){\LineV{#2}} \put(#1,0){\LineV{#2}}
\put(0,0){\line(1,1){#2}}\put(0,#2){\line(1,-1){#2}}
\put(#1,0){\line(-1,1){#2}}\put(#1,#2){\line(-1,-1){#2}}
\put(0,0.3){\line(1,1){0.3}}\put(1.5,0.3){\line(-1,1){0.3}}
\put(0,0){\circl}\put(0,#2){\circl}
\put(#1,0){\circl}\put(#1,#2){\circl}}

\newcommand{\SEboxxP}
{\put(0,-0.3){\LineH{2}} \put(0,0.3){\LineH{2}}
\put(0,-0.3){\LineV{0.6}} \put(2,-0.3){\LineV{0.6}}
\put(0,-0.0){\makebox(0,0){\multiput(0,0)(0,0.2){2}{\line(1,0){2}}}}
\put(0.25,0){\makebox(0,0){\multiput(0,0)(0.3,0){6}{\line(0,1){0.6}}}}
\put(0,-0.3){\circl}\put(0,0.3){\circl}
\put(2,-0.3){\circl}\put(2,0.3){\circl}}

\newcommand{\VQED}{\Photon}

\newcommand{\VQEDT}{\PhotonT}

\newcommand{\SEboxx}
{\put(0,-0.3){\LineH{2}}\put(0,0.3){\LineH{2}}
\put(0,-0.4){\multiput(0,0)(0.10,0){20}{$\cdot$}}
\put(0,-0.1){\multiput(0,0)(0.10,0){20}{$\cdot$}}
\put(0,-0.25){\multiput(0,0)(0.10,0){20}{$\cdot$}}
\put(0,0.05){\multiput(0,0)(0.10,0){20}{$\cdot$}}
\put(0,-0.3){\circl}\put(0,0.3){\circl}
\put(2,-0.3){\circl}\put(2,0.3){\circl}}

\newcommand{\SEboxxx}
{\put(0,-0.3){\LineH{3}} \put(0,0.3){\LineH{3}}
\put(0,-0.3){\LineV{0.6}} \put(3,-0.3){\LineV{0.6}}
\put(0,-0.4){\multiput(0,0)(0.10,0){30}{$\cdot$}}
\put(0,-0.1){\multiput(0,0)(0.10,0){30}{$\cdot$}}
\put(0,-0.25){\multiput(0,0)(0.10,0){30}{$\cdot$}}
\put(0,0.05){\multiput(0,0)(0.10,0){30}{$\cdot$}}
\put(0,-0.3){\circl}\put(0,0.3){\circl}
\put(3,-0.3){\circl}\put(3,0.3){\circl}}

\newcommand{\SEbox}
{\put(0,0){\makebox(0,0.35){\EEbox{1}{0.7}}} }

\newcommand{\EboxG}
{\put(0,0){\makebox(0,0.7){\EEbox{2}{0.7}}}
\put(0,0){\circl}\put(2,0){\circl}
\put(0,0.7){\circl}\put(2,0.7){\circl}}

\newcommand{\SEboxP}
{\put(0,0){\makebox(0,0){\multiput(0,0)(0,0.2){3}{\line(1,0){1}}}}
\put(0,0){\makebox(0,0){\multiput(0,0)(0.2,0){5}{\line(0,1){0.7}}}}
  \put(0,0){\makebox(0,0){\Ebox{1}{0.7}}}}

\newcommand{\photonG}
{\qbezier(0,0)(0.08333,0.125)(0.1666667,0)
\qbezier(0.1666667,0)(0.25,-0.125)(0.3333333,0)
\qbezier(0.3333333,0)(0.416667,0.125)(0.5,0)

\qbezier(0.5,0)(0.583333,-0.125)(0.666667,0)
\qbezier(0.666667,0)(0.75,0.125)(0.833333,0)
\qbezier(0.833333,0)(0.916667,-0.125)(1,0)

\qbezier(1,0)(1.083333,0.125)(1.166667,0)
\qbezier(1.166667,0)(1.25,-0.125)(1.333333,0)
\qbezier(1.333333,0)(1.416667,0.125)(1.5,0)

\qbezier(1.5,0)(1.583333,-0.125)(1.666667,0)
\qbezier(1.666667,0)(1.75,0.125)(1.833333,0)
\qbezier(1.833333,0)(1.916667,-0.125)(2,0)
\put(0,0){\circle*{0.2}}\put(2,0){\circle*{0.2}} }

\newcommand{\photonTG}[3]
{\put(0,0){\photonn{}{#2}{}}\put(1,0){\photonn{}{#2}{}}}

\newcommand{\photong}
{\qbezier(0,0)(0.08333,0.125)(0.1666667,0)
\qbezier(0.1666667,0)(0.25,-0.125)(0.3333333,0)
\qbezier(0.3333333,0)(0.416667,0.125)(0.5,0)

\qbezier(0.5,0)(0.583333,-0.125)(0.666667,0)
\qbezier(0.666667,0)(0.75,0.125)(0.833333,0)
\qbezier(0.833333,0)(0.916667,-0.125)(1,0)

\qbezier(1,0)(1.083333,0.125)(1.166667,0)
\qbezier(1.166667,0)(1.25,-0.125)(1.333333,0)
\qbezier(1.333333,0)(1.416667,0.125)(1.5,0)
\put(0,0){\circle*{0.20}}}

\newcommand{\photonnn}
{\qbezier(0,0)(0.08333,0.125)(0.1666667,0)
\qbezier(0.1666667,0)(0.25,-0.125)(0.3333333,0)
\qbezier(0.3333333,0)(0.416667,0.125)(0.5,0)

\qbezier(0.5,0)(0.583333,-0.125)(0.666667,0)
\qbezier(0.666667,0)(0.75,0.125)(0.833333,0)
\qbezier(0.833333,0)(0.916667,-0.125)(1,0)

\qbezier(1,0)(1.083333,0.125)(1.166667,0)
\qbezier(1.166667,0)(1.25,-0.125)(1.333333,0)
\qbezier(1.333333,0)(1.416667,0.125)(1.5,0) }

\newcommand{\photonGENE}[3]
{\qbezier(0,0)(0.17,-0.04)(0.2,0.1)
\qbezier(0.2,0.1)(0.23,0.32)(0.4,0.2)
\qbezier(0.4,0.2)(0.57,0.16)(0.6,0.3)
\qbezier(0.6,0.3)(0.63,0.52)(0.8,0.4)
\qbezier(0.8,0.4)(0.97,0.36)(1,0.5)
\qbezier(1,0.5)(1.03,0.72)(1.2,0.6)
\qbezier(1.2,0.6)(1.37,0.56)(1.4,0.7)
\qbezier(1.4,0.7)(1.43,0.92)(1.6,0.8)
\qbezier(1.6,0.8)(1.77,0.76)(1.8,0.9)
\qbezier(1.8,0.9)(1.83,1.12)(2,1)
\put(0,0){\circle*{0.20}} \put(2,1){\circle*{0.20}}
\put(1,0.85){\makebox(0,-0){$#1$}}
\put(-0.35,-1){\makebox(0,2){$#2$}}
\put(2.35,0){\makebox(0,2){$#3$}}}

\newcommand{\photonGWNW}[3]
{\qbezier(0,0)(-0.17,-0.04)(-0.2,0.1)
\qbezier(-0.2,0.1)(-0.23,0.32)(-0.4,0.2)
\qbezier(-0.4,0.2)(-0.57,0.16)(-0.6,0.3)
\qbezier(-0.6,0.3)(-0.63,0.52)(-0.8,0.4)
\qbezier(-0.8,0.4)(-0.97,0.36)(-1,0.5)
\qbezier(-1,0.5)(-1.03,0.72)(-1.2,0.6)
\qbezier(-1.2,0.6)(-1.37,0.56)(-1.4,0.7)
\qbezier(-1.4,0.7)(-1.43,0.92)(-1.6,0.8)
\qbezier(-1.6,0.8)(-1.77,0.76)(-1.8,0.9)
\qbezier(-1.8,0.9)(-1.83,1.12)(-2,1)
\put(0,0){\circle*{0.2}} \put(-2,1){\circle*{0.2}}
\put(-1,0.5){\makebox(0.6,0.4){$#1$}}
\put(0.35,-1){\makebox(0,2){$#3$}}
\put(-2.35,0){\makebox(0,2){$#2$}}}

\newcommand{\photonHG}
{\qbezier(0,0)(0.08333,0.125)(0.1666667,0)
\qbezier(0.1666667,0)(0.25,-0.125)(0.3333333,0)
\qbezier(0.3333333,0)(0.416667,0.125)(0.5,0)

\qbezier(0.5,0)(0.583333,-0.125)(0.666667,0)
\qbezier(0.666667,0)(0.75,0.125)(0.833333,0)
\qbezier(0.833333,0)(0.916667,-0.125)(1,0)

\qbezier(1,0)(1.083333,0.125)(1.166667,0)
\qbezier(1.166667,0)(1.25,-0.125)(1.333333,0)
\qbezier(1.333333,0)(1.416667,0.125)(1.5,0)
\put(0,0){\circle*{0.20}}\put(1.5,0){\circle*{0.20}} }

\renewcommand{\photonH}
{\photonHG\put(0.75,-0.1){\VectorR}}

\newcommand{\photonhp}
{\qbezier(0,0)(0.08333,0.125)(0.1666667,0)
\qbezier(0.1666667,0)(0.25,-0.125)(0.3333333,0)
\qbezier(0.3333333,0)(0.416667,0.125)(0.5,0)

\qbezier(0.5,0)(0.583333,-0.125)(0.666667,0)
\qbezier(0.666667,0)(0.75,0.125)(0.833333,0)
\qbezier(0.833333,0)(0.916667,-0.125)(1,0)
\put(0,0){\circle*{0.20}} \put(0.5,0){\VectorR}}

\newcommand{\photonhpL}
{\qbezier(0,0)(0.08333,0.125)(0.1666667,0)
\qbezier(0.1666667,0)(0.25,-0.125)(0.3333333,0)
\qbezier(0.3333333,0)(0.416667,0.125)(0.5,0)

\qbezier(0.5,0)(0.583333,-0.125)(0.666667,0)
\qbezier(0.666667,0)(0.75,0.125)(0.833333,0)
\qbezier(0.833333,0)(0.916667,-0.125)(1,0)
\put(1,0){\circle*{0.20}} \put(0.5,0){\Vector}}

\newcommand{\photonhpG}
{\qbezier(0,0)(0.08333,0.125)(0.1666667,0)
\qbezier(0.1666667,0)(0.25,-0.125)(0.3333333,0)
\qbezier(0.3333333,0)(0.416667,0.125)(0.5,0)

\qbezier(0.5,0)(0.583333,-0.125)(0.666667,0)
\qbezier(0.666667,0)(0.75,0.125)(0.833333,0)
\qbezier(0.833333,0)(0.916667,-0.125)(1,0)
\put(0,0){\circle*{0.20}} }

\newcommand{\photonh}
{\put(0,0){\photonhp} \put(1,0){\circle*{0.20}}}

\newcommand{\photonhG}
{\put(0,0){\photonhpG} \put(1,0){\circle*{0.20}}}

\newcommand{\photonV}[0]
{\qbezier(0,0)(-0.1666667,0.111111)(0,0.222222)
\qbezier(0,0.222222)(0.1666667,0.333333)(0,0.444444)
\qbezier(0,0.444444)(-0.1666667,0.55555)(0,0.666667)
\qbezier(0,0.666667)(0.1666667,0.777777)(0,0.888889)
\qbezier(0,0.888889)(-0.1666667,1)(0,1.111111)
\put(0,0){\circle*{0.20}} \put(0,1.111){\circle*{0.20}} }

\newcommand{\photonNEG}[3]
{\qbezier(0,0)(0.22,-0.02)(0.2,0.2)
\qbezier(0.2,0.2)(0.18,0.42)(0.4,0.4)
\qbezier(0.4,0.4)(0.62,0.38)(0.6,0.6)
\qbezier(0.6,0.6)(0.58,0.82)(0.8,0.8)
\qbezier(0.8,0.8)(1.02,0.78)(1,1)
\qbezier(1,1)(0.98,1.22)(1.2,1.2)
\qbezier(1.2,1.2)(1.42,1.18)(1.4,1.4)
\qbezier(1.4,1.4)(1.38,1.62)(1.6,1.6)
\qbezier(1.6,1.6)(1.82,1.58)(1.8,1.8)
\qbezier(1.8,1.8)(1.78,2.02)(2,2)\put(1,1){\makebox(-0.6,0.4){$#1$}}
\put(-0.35,-1){\makebox(0,2){$#2$}}
\put(2.35,1){\makebox(0,2){$#3$}}
\put(0,0){\circle*{0.20}}\put(2,2){\circle*{0.20}}
 }

\newcommand{\photonNWG}[3]
{\qbezier(0,0)(-0.22,-0.02)(-0.2,0.2)
\qbezier(-0.2,0.2)(-0.18,0.42)(-0.4,0.4)
\qbezier(-0.4,0.4)(-0.62,0.38)(-0.6,0.6)
\qbezier(-0.6,0.6)(-0.58,0.82)(-0.8,0.8)
\qbezier(-0.8,0.8)(-1.02,0.78)(-1,1) \qbezier(-1,1)
(-0.98,1.22)(-1.2,1.2) \qbezier(-1.2,1.2)(-1.42,1.18)(-1.4,1.4)
\qbezier(-1.4,1.4)(-1.38,1.62)(-1.6,1.6)
\qbezier(-1.6,1.6)(-1.82,1.58)(-1.8,1.8)
\qbezier(-1.8,1.8)(-1.78,2.02)(-2,2)
\put(-1,1){\makebox(0.4,0.7){$#1$}}
\put(-2.35,2){\makebox(0,0){$#2$}}
\put(0.35,0){\makebox(0,0){$#3$}}
}

\newcommand{\photonENEG}[3]
{\qbezier(0,0)(0.17,-0.04)(0.2,0.1)
\qbezier(0.2,0.1)(0.23,0.32)(0.4,0.2)
\qbezier(0.4,0.2)(0.57,0.16)(0.6,0.3)
\qbezier(0.6,0.3)(0.63,0.52)(0.8,0.4)
\qbezier(0.8,0.4)(0.97,0.36)(1,0.5)
\qbezier(1,0.5)(1.03,0.72)(1.2,0.6)
\qbezier(1.2,0.6)(1.37,0.56)(1.4,0.7)
\qbezier(1.4,0.7)(1.43,0.92)(1.6,0.8)
\qbezier(1.6,0.8)(1.77,0.76)(1.8,0.9)
\qbezier(1.8,0.9)(1.83,1.12)(2,1)
\put(1,0.85){\makebox(0,-0){$#1$}}
\put(-0.35,-1){\makebox(0,2){$#2$}}
\put(2.35,0){\makebox(0,2){$#3$}} \put(0,0){\circle*{0.20}}
\put(2,1){\circle*{0.20}}}

\newcommand{\photonWNWG}[3]
{\qbezier(0,0)(-0.17,-0.04)(-0.2,0.1)
\qbezier(-0.2,0.1)(-0.23,0.32)(-0.4,0.2)
\qbezier(-0.4,0.2)(-0.57,0.16)(-0.6,0.3)
\qbezier(-0.6,0.3)(-0.63,0.52)(-0.8,0.4)
\qbezier(-0.8,0.4)(-0.97,0.36)(-1,0.5)
\qbezier(-1,0.5)(-1.03,0.72)(-1.2,0.6)
\qbezier(-1.2,0.6)(-1.37,0.56)(-1.4,0.7)
\qbezier(-1.4,0.7)(-1.43,0.92)(-1.6,0.8)
\qbezier(-1.6,0.8)(-1.77,0.76)(-1.8,0.9)
\qbezier(-1.8,0.9)(-1.83,1.12)(-2,1) \put(0,0){\circle*{0.15}}
\put(-2,1){\circle*{0.15}} \put(-1,0.5){\makebox(0.6,0.4){$#1$}}
\put(0.35,-1){\makebox(0,2){$#3$}}
\put(-2.35,0){\makebox(0,2){$#2$}}}

\newcommand{\CrossphotonsG}[6]
{\put(0,0){\photonNEG{#5}{}{}} \put(2,0){\photonNWG{#6}{}{}}
\put(-0.35,0){\makebox(0,0){$#1$}}
\put(2.35,2){\makebox(0,0){$#2$}}
\put(2.35,0){\makebox(0,0){$#3$}}
\put(-0.35,2){\makebox(0,0){$#4$}}
 \put(0,0){\circle*{0.20}} \put(2,0){\circle*{0.20}}
 \put(0,2){\circle*{0.20}} \put(2,2 ){\circle*{0.20}}}

\newcommand{\photonHSG}[4]
{\qbezier(0,0)(0.08333,0.125)(0.1666667,0)
\qbezier(0.1666667,0)(0.25,-0.125)(0.3333333,0)
\qbezier(0.3333333,0)(0.416667,0.125)(0.5,0)
\qbezier(0.5,0)(0.583333,-0.125)(0.666667,0)
\qbezier(0.666667,0)(0.75,0.125)(0.833333,0)
\qbezier(0.833333,0)(0.916667,-0.125)(1,0)
\put(0.5,0.35){\makebox(0,0){$#1$}} \put(0,0){\circle*{0.20}}
\put(1,0){\circle*{0.20}}
\put(0,-0.5){\makebox(0,0){#2}}\put(1,-0.5){\makebox(0,0){#3}}}

\newcommand{\PotintG}
{\linethickness{0.3mm}\put(0,0)\dashH
\put(1.35,0){\makebox(0,0){\large$\times$}}
\put(0,0){\circle*{0.20}} }

\newcommand{\EffpotG}
{\linethickness{0.3mm}\put(0,0){\multiput(0.05,0)(0.25,0){5}{\line(1,0){0.15}}}
\put(1.55,0){\makebox(0,0){\large$\times$}}
\put(1.55,0){\circle{0.5}} \put(0,0){\circle*{0.20}} }

\newcommand{\EffpotGG}
{\linethickness{0.3mm}\put(0,0){\multiput(0.05,0)(0.25,0){5}{\line(1,0){0.15}}}
\put(1.6,0){\makebox(0,0){\large$\times$}}
\put(1.6,0){\circle{0.5}} \put(1.6,0){\circle{0.7}}
\put(0,0){\circle*{0.20}} }

\newcommand{\elstatG}[3]
{\multiput(0.06,0)(0.25,0){8}{\line(1,0){0.15}}
\put(1,0.35){\makebox(0,0){$#1$}} \put(0,0){\circle*{0.2}}
\put(2,0){\circle*{0.2}} \put(-0.35,0){\makebox(0,0){#2}}
\put(2.35,0){\makebox(0,0){#3}}}

\newcommand{\elstatHG}[3]
{\linethickness{0.3mm}\multiput(0.06,0)(0.25,0){6}{\line(1,0){0.15}}
\put(0.75,0.25){\makebox(0,0){$#1$}} \put(0,0){\circle*{0.2}}
\put(1.5,0){\circle*{0.2}} \put(-0.35,0){\makebox(0,0){#2}}
\put(2.35,0){\makebox(0,0){#3}}}

\newcommand{\VPloopG}[1]
{\put(0,0){\circle{1}}\put(0.025,-0.025){\circle{1}}\put(0.025,-0.025){\circle{1}}
\put(0.025,-0){\circle{1}}\put(0,-0.025){\circle{1}}
\put(0.85,0){\makebox(0,0){$#1$}}}

\newcommand{\LoopG}[2]
{\put(0,0){\Oval{0.6}{1.25}} \put(-0.3,0){\VectorDn}
\put(-0.65,0){\makebox(0,0){$#1$}}
\put(0.65,0){\makebox(0,0){$#2$}}}

\newcommand{\photonSEG}[5]
{\put(0,0){\photonHSG{#3}{#4}{}{}} \put(1.5,0){\VPloopD{#1}{#2}}
\put(2,0){\photonHSG{#5}{}{}{}} }

\newcommand{\GG}[2]
{\put(0,0){\LineV{#1}} \put(0,0){\circl}
\put(0,#1){\circl}\put(0,#2){\VectorUp} }

\newcommand{\Gn}[1]
{\put(0,0){\LineV{#1}} \put(0,0){\circl} \put(0,#1){\circl} }

\newcommand{\Gnn}[1]
{\put(0,0){\LineV{#1}} \put(0,#1){\circl} }

\newcommand{\GGD}[3]
{\put(0,0){\LineV{#1}} \put(0,0){\circl}
\put(0,#1){\circl}\put(0,#2){\VectorUp} \put(0,#3){\VectorUp}}

\newcommand{\GGT}[4]
{\put(0,0){\LineV{#1}} \put(0,0){\circl}
\put(0,#1){\circl}\put(0,#2){\VectorUp}
\put(0,#3){\VectorUp}\put(0,#4){\VectorUp}}

\newcommand{\GGF}[5]
{\put(0,0){\LineV{#1}} \put(0,0){\circl}
\put(0,#1){\circl}\put(0,#2){\VectorUp}
\put(0,#3){\VectorUp}\put(0,#4){\VectorUp}\put(0,#5){\VectorUp}}

\newcommand{\GGFm}[6]
{\put(0,0){\LineV{#1}} \put(0,0){\circl}
\put(0,#1){\circl}\put(0,#2){\VectorUp}
\put(0,#3){\VectorUp}\put(0,#4){\VectorUp}\put(0,#5){\VectorUp}
\put(0,#6){\VectorUp}}

\newcommand{\Gg}[1]
{\put(0,0){\LineV{#1}} \put(0,0){\circle} \put(0,#1){\circl}}

\newcommand{\GGC}[5]
{\put(0,0){\LineV{#1}} \put(0,#3){\circl}\put(0,#4){\circl}
}

\newcommand{\GGc}[3]
{\put(0,0){\LineV{#1}} \put(0,#3){\circl}
}

\newcommand{\GGUp}[0]
  {\put(0,0){\LineV{1.5}}
 }

\newcommand{\GGUpR}[0]
  {\put(0,0){\line(1,3){0.5}}
  \put(0,0){\vector(1,3){0.30}}
  }

\newcommand{\GGDnL}[0]
  {\put(0,0){\line(-1,3){0.5}}
  \put(0,0){\vector(-1,3){0.30}}
   }

\newcommand{\GGDnR}[0]
  {\put(0,0){\line(1,-3){0.5}}
  \put(0,0){\vector(1,-3){0.30}}
   }

\newcommand{\GGG}[2]
{\linethickness{0.8mm}\put(0,0){\line(0,1){#1}} \put(0,0){\circl}
\put(0,#1){\circl}\put(0.035,#2)
{\VectorUp}\put(-0.035,#2){\VectorUp} }

\newcommand{\Pot}[0]
{ \put(0,0){\photonnH}  \put(1.2,-0.15){\Large$\times$}
\put(0,0){\circl}}

\newcommand{\PotL}[0]
{ \put(0.5,0){\photonH}  \put(1.7,-0.2){\Large$\times$}
\put(-0.475,0){\VectorUp} \put(0,0){\circle{1}}}

\newcommand{\VP}[0]
{\put(0,0){\Circle{1}} 
 \put(0.1,0.5){\vector(1,0){0}}
 }

\newcommand{\VPG}[0]
{\put(2,0){\LoopT{1}} \put(0,0){\photonH}
\put(2.45,0.05){\VectorUp} }

\newcommand{\VPh}[0]
{\put(1.5,0){\LoopT{1}} \put(0,0){\photonh}
  \put(1.95,0.05){\VectorUp}}

\newcommand{\VPhh}[0]
{\put(1.5,0){\LoopT{1}} \put(0,0){\photonh}
 \put(1.5,0.475){\Vector}\put(1.5,-0.475){\VectorR}
  }

\newcommand{\VPLG}[0]
{\put(-2,0){\LoopT{1}} \put(-1.5,0){\photonH{}{}{}}
\put(-2.5,-0.05){\VectorUp}
 }

\newcommand{\VPLh}[0]
{\put(-1.5,0){\Circle{1}} \put(-1,0){\photonh}
\put(-2,-0.05){\VectorUp}
 }

\newcommand{\VPPG}[0]
{\put(0,0){\LoopTh{1.5}} \put(-0.75,0){\photonH}
\put(0,0.70){\Vector}\put(0,-0.70){\VectorR} }

\newcommand{\VPPGLab}[2]
{\put(0,0){\LoopTh{1.5}} \put(-0.75,0){\photonH}
\put(0,1.1){\makebox(0,0){$#1$}}\put(0,-1,1){\makebox(0,0){$#2$}}
\put(0,0.70){\Vector}\put(0,-0.70){\VectorR }}

\newcommand{\IrrPot}[0]
{\put(-0.1,0){\makebox(0,0){\multiput(0,0)(0.3,0){7}{$\times$}}}
 \put(0,0){\circle*{0.25}} \put(2,0){\circle*{0.25}}}

\newcommand{\IrrPots}[0]
{\put(-0.1,0){\makebox(0,0){\multiput(0,0)(0.3,0){5}{$\times$}}}
 \put(0,0){\circle*{0.15}} \put(1.5,0){\circle*{0.15}}}

\newcommand{\IrrPotS}[0]
{\put(-0.1,0){\makebox(0,0){\multiput(0,0)(0.3,0){4}{$\times$}}}
 \put(0,0){\circle*{0.25}}}

\newcommand{\LineSG}[1]
{\linethickness{0.75mm} \put(0,0){\line(1,0){#1}}
\put(0,0){\circle*{0.25}} \put(#1,0){\circle*{0.25}}}

\newcommand{\LineSS}[1]
{\linethickness{1mm} \put(0,0){\line(1,0){#1}}
\put(0,0){\circle*{0.25}} }

\newcommand{\Handle}[0]
{\put(0,0){\LoopT{1}} \put(0.5,0){\photonH}\put(2.5,0){\LoopT{1}}
\put(-0.47,-0.025){\VectorUp}\put(2.97,-0.025){\VectorUp} }

\newcommand{\Circle}[1]
{\thicklines\put(0.,0){\circle{#1}}}

\newcommand{\HandleLab}[2]
{\put(0,0){\Circle{1}}
\put(0.5,0){\photon{}{}{}}\put(3,0){\Circle{1}}
\put(-0.8,0){\makebox(0,0){$#1$}}\put(3.8,0){\makebox(0,0){$#2$}}
\put(-0.475,0){\VectorUp}\put(3.475,0){\VectorUp}}

\newcommand{\PhInt}[0]
{\put(0,0){\photon{}{}{}}\put(0,0){\vector(-1,1){0.6}}\put(2,0){\vector(1,1){0.6}}
 \put(-0.6,-0.6){\vector(1,1){0.6}}\put(2.6,-0.6){\vector(-1,1){0.6}}}

\newcommand{\PhIntS}[0]
{\put(0,0){\photong}\put(0,0){\vector(-1,1){0.6}}\put(2,0){\vector(1,1){0.6}}
 \put(-0.6,-0.6){\vector(1,1){0.6}}\put(2.6,-0.6){\vector(-1,1){0.6}}}

\newcommand{\PhIntLab}[4]
 {\put(0,0){\photon{}{}{}}\put(0,0){\vector(-1,1){0.6}}\put(2,0){\vector(1,1){0.6}}
 \put(-0.6,-0.6){\vector(1,1){0.6}}\put(2.6,-0.6){\vector(-1,1){0.6}}
 \put(-1,0.8){\makebox(0,0){#1}}\put(3,0.8){\makebox(0,0){#2}}
 \put(-1,-0.8){\makebox(0,0){#3}}\put(3,-0.8){\makebox(0,0){#4}}
 \put(0,0.3){1}\put(1.75,0.3){2}}

\newcommand{\ElSEG}[0]
{\qbezier(0,-1)(.2025,-1.1489)(0.3420,-0.9397)
\qbezier(0.3420,-0.9397)(0.4167,-0.7217)(0.6428,-0.766)
\qbezier(0.6428,-0.766)(0.8937,-0.7499)(0.866,-0.5)
\qbezier(0.866,-0.5)(0.7831,-0.2850)(0.9848,-0.1736)
\qbezier(0.9848,-0.1736)(1.1667,0)(0.9848,0.1736)
\qbezier(0.9848,0.1736)(0.7831,0.2850)(0.866,0.5)
\qbezier(0.866,0.5)(0.8937,0.7499)(0.6428,0.766)
\qbezier(0.6428,0.766)(0.4167,0.7217)(0.3420,0.9397)
\qbezier(0.3420,0.9397)(.2025,1.1489)(0,1)
\put(0,-1){\circle*{0.20}}\put(0,1){\circle*{0.20}} }

\newcommand{\ElSEg}
{\setlength{\unitlength}{0.4cm}\qbezier(0,-1)(.2025,-1.1489)(0.3420,-0.9397)
\qbezier(0.3420,-0.9397)(0.4167,-0.7217)(0.6428,-0.766)
\qbezier(0.6428,-0.766)(0.8937,-0.7499)(0.866,-0.5)
\qbezier(0.866,-0.5)(0.7831,-0.2850)(0.9848,-0.1736)
\qbezier(0.9848,-0.1736)(1.1667,0)(0.9848,0.1736)
\qbezier(0.9848,0.1736)(0.7831,0.2850)(0.866,0.5)
\qbezier(0.866,0.5)(0.8937,0.7499)(0.6428,0.766)
\qbezier(0.6428,0.766)(0.4167,0.7217)(0.3420,0.9397)
\qbezier(0.3420,0.9397)(.2025,1.1489)(0,1)
\put(0,-1){\circle*{0.225}}\put(0,1){\circle*{0.225}}}

\newcommand{\ElSEgg}
{\setlength{\unitlength}{0.8cm}\qbezier(0,-1)(.2025,-1.1489)(0.3420,-0.9397)
\qbezier(0.3420,-0.9397)(0.4167,-0.7217)(0.6428,-0.766)
\qbezier(0.6428,-0.766)(0.8937,-0.7499)(0.866,-0.5)
\qbezier(0.866,-0.5)(0.7831,-0.2850)(0.9848,-0.1736)
\qbezier(0.9848,-0.1736)(1.1667,0)(0.9848,0.1736)
\qbezier(0.9848,0.1736)(0.7831,0.2850)(0.866,0.5)
\qbezier(0.866,0.5)(0.8937,0.7499)(0.6428,0.766)
\qbezier(0.6428,0.766)(0.4167,0.7217)(0.3420,0.9397)
\qbezier(0.3420,0.9397)(.2025,1.1489)(0,1)
\put(0,-1){\circle*{0.11}}\put(0,1){\circle*{0.11}}}

\newcommand{\ElSELG}[0]
{\qbezier(0,-1)(-.2025,-1.1489)(-0.3420,-0.9397)
\qbezier(-0.3420,-0.9397)(-0.4167,-0.7217)(-0.6428,-0.766)
\qbezier(-0.6428,-0.766)(-0.8937,-0.7499)(-0.866,-0.5)
\qbezier(-0.866,-0.5)(-0.7831,-0.2850)(-0.9848,-0.1736)
\qbezier(-0.9848,-0.1736)(-1.1667,0)(-0.9848,0.1736)
\qbezier(-0.9848,0.1736)(-0.7831,0.2850)(-0.866,0.5)
\qbezier(-0.866,0.5)(-0.8937,0.7499)(-0.6428,0.766)
\qbezier(-0.6428,0.766)(-0.4167,0.7217)(-0.3420,0.9397)
\qbezier(-0.3420,0.9397)(-.2025,1.1489)(0,1)
\put(0,-1){\circle*{0.20}}\put(0,1){\circle*{0.20}} }